\documentclass[useAMS,usenatbib]{mn2e}
\usepackage{natbib, graphicx, times}
\usepackage{amsmath}
\usepackage{mathtools}
\usepackage{epstopdf}
\usepackage{amssymb}
\usepackage{extarrows}
\usepackage{color}
\usepackage{enumitem}
\usepackage{scrextend}
\newcommand{\apj}{ApJ}
\newcommand{\apjl}{ApJ}

\newcommand{\aap}{A \& A}
\newcommand{\aaps}{A \& AS}

\newcommand{\icarus}{Icarus}
\newcommand{\mnras}{MNRAS}

\newcommand{\nat}{Nature}

\setlist{leftmargin=7.5mm}
\newcommand{\tab}[1]{\hspace{.02\textwidth}\rlap{#1}}

\newcommand{\lsim}{\mathrel{\rlap{\lower4pt\hbox{\hskip1pt$\sim$}}
    \raise1pt\hbox{$<$}}}                % less than or approx. symbol
\newcommand{\gsim}{\mathrel{\rlap{\lower4pt\hbox{\hskip1pt$\sim$}}
    \raise1pt\hbox{$>$}}}                % greater than or approx. symbol
   
\newcommand{\cbf}{} % temporary bold face comments

\def\arcsec{\hbox{$^{\hbox{\rlap{\hbox{\lower4pt\hbox{$\,\prime\prime$}}}\hbox{$\frown$}}}$}}

\title[Images of growing planet-induced vortices]{Observational diagnostics of elongated planet-induced vortices with realistic planet formation timescales}

\author[Hammer, Pinilla, Kratter, \& Lin]{Michael Hammer$^{1}$\thanks{E-mail: mhammer@as.arizona.edu},
Paola Pinilla$^{1,3}$,
Kaitlin M. Kratter$^{1}$,
and Min-Kai Lin$^{2}$\\
$^{1}$ Steward Observatory, University of Arizona, Tucson, AZ 85721, USA \\
$^{2}$ Institute of Astronomy and Astrophysics, Academia Sinica, Taipei 10617, Taiwan \\
$^{3}$ Hubble Fellow
}

\begin{document}

\date{Accepted XXX. Received YYY; in original form ZZZ}

\pagerange{\pageref{firstpage}--\pageref{lastpage}} \pubyear{2016}

\maketitle

\label{firstpage}

\begin{abstract}
Gap-opening planets can generate dust-trapping vortices that may explain some of the latest discoveries of high-contrast crescent-shaped dust asymmetries in transition discs. While planet-induced vortices were previously thought to have concentrated shapes, recent computational work has shown that these features naturally become much more elongated in the gas when simulations account for the relatively long timescale over which planets accrete their mass. In this work, we conduct two-fluid hydrodynamical simulations of vortices induced by slowly-growing Jupiter-mass planets in discs with very low viscosity ($\alpha = 3 \times 10^{-5}$). We simulate the dust dynamics for {\cbf four particle sizes spanning 0.3 mm to 1 cm in order to produce synthetic ALMA images. In our simulations, we find that an elongated vortex still traps dust, but not directly at its center. With a flatter pressure bump and disruptions from the planet's overlapping spiral density waves, the dust instead circulates around the vortex. This motion (1) typically carries the peak off-center, (2) spreads the dust out over a wider azimuthal extent $\geq 180^{\circ}$, (3) skews the azimuthal profile towards the front of the vortex, and (4) can also create double peaks in newly-formed vortices. In particular, we expect that the most defining observational signature, a peak offset of more than $30^{\circ}$, should be detectable $>30\%$ of the time in observations with a beam diameter of at most the planet's separation from its star.
}

\end{abstract}


%@arxiver{azimuthalDensityOnePanelComparison_0300-0800.pdf,intensityCartGridTaperComparison_0300-0800.png}

% #### INTRODUCTION #### %

\section{Introduction} \label{sec:intro}

In the last few years, a combination of individual observations and surveys using the Atacama Large Millimeter Array (ALMA) have imaged a large sample of transition discs, many of which have large-scale asymmetries in the dust at mm and sub-mm wavelengths \citep[e.g.][]{vanDerMarel13, casassus13, perez14, ansdell16, pascucci16, barenfeld16}. One possible explanation for these features is that they are dust-trapping vortices induced by giant planets that triggered the Rossby Wave instability \citep[RWI:][]{lovelace99, li00, li01} as they opened up gaps in low-viscosity discs \citep[e.g.][]{li05, deValBorro07}.
	
The idea that these asymmetries are signatures of planets can be tested in multiple ways. Observational studies have attempted to validate these features as vortices by showing that the dust in these asymmetries becomes more concentrated at larger grain sizes \citep{vanDerMarel15b}. Meanwhile, computational studies have also tried to support this explanation by showing that synthetic images generated from simulations of a vortex's dust distribution closely match the images taken by ALMA \citep{bae16}. Nonetheless, even if an asymmetry is a vortex, it may not have been created by a planet. The build-up of gas at a dead zone boundary can also generate a vortex \citep[e.g.][]{regaly12, lyra12, miranda17}. Furthermore, alternate ideas such as the presence of a close-in companion star adding eccentricity to the disc \citep[][]{ataiee13, ragusa17, price18} are also possible. These scenarios that do not rely on planets may still prove to be better explanations for some, if not all of these asymmetries.

%alternate ideas such as the presence of a close-in companion star adding eccentricity to the disc \citep[][]{ragusa17} or a vortex induced by a dead zone boundary \citep[e.g.][]{regaly12, lyra12, miranda17} may still prove to be better explanations for some, if not all of these asymmetries.
	
To understand if disc asymmetries can be evidence for undetected planets, it is useful to simulate the dynamics of dust in a disc in order to create synthetic images at mm wavelengths -- the range of emission that traces particle sizes expected to be trapped by vortices in the outer part of a disk. Yet, previous efforts to simulate images of these vortices have neglected the time it takes for a planet to grow to Jupiter-size, a key parameter in determining the properties of a vortex \citep{hammer17}. Computational models of core accretion --- the preferred method for forming most gas giants --- show that a giant planet will obtain most of its mass during its runaway growth phase, which can last for thousands of orbits or longer in discs with the low viscosities needed for vortices to form \citep{pollack96, lissauer09}. Taking these timescales into account, \cite{hammer17} found that a giant planet can first trigger the RWI while it is still in its growth phase. At this point, it will create a weaker, elongated vortex compared to the stronger, more concentrated vortices induced by planets grown in less than 100 orbits. With these differences in the shape of the gas vortex, \cite{hammer17} expect the dust component seen in observations to appear differently as well.
	
In this work, we investigate the impact of a gas vortex having an elongated structure on dust-trapping. We generate synthetic images of both elongated vortices and concentrated vortices in order to identify unique observational signatures that can be used to differentiate these two cases.

%We focus our study on a system with a planet-to-star mass ratio of 10$^{-3}$, which is equivalent to a Jupiter-mass planet around a solar mass star.
	
The paper is organized as follows: In Section~\ref{sec:methods}, we describe our two-fluid (gas and dust) 2-D hydrodynamical simulations and our method for generating synthetic images. In Section~\ref{sec:results}, we compare the two classes of vortices in two-fluid simulations and synthetic dust continuum images. We propose several criteria needed to classify an observed asymmetry as an elongated vortex. In Section~\ref{sec:applications}, we apply our results to existing and future observations. In Section~\ref{sec:conclusions}, we summarize our results. \\

%%Outline: \\
%%(1) We observe transition discs with large asymmetries. \\
%%(1b) Cite transition discs. \\
%%(2) Vortices generated by the RWI can explain these asymmetries. \\
%%(2b) Cite planet-induced vortex papers, including support for this explanation over others. \\
%%(3) Recently, it has been shown that if a planet triggers the RWI while it is still growing, it will create a weaker vortex that is much more than those produced by a rapidly-grown planet. \\
%%(4) Actual observations of disks with vortex candidates are of the dust, with ALMA. \\
%%(5) Investigate whether differences in gas structure carry over to dust structure \\
%%(6) Simulate dust structure and radiative transfer \\
%%(7) We show that the azimuthal extent, as well flatness (lack of contours) of the elongated vortices are the key features for identifying and differentiating them from concentrated vortices. \\
%%(8) In Section 2, methods. In Section 3, results (dust structure and radiative transfer images). In Section 4, discussion.

% #### METHODS #### %

\section{Methods} \label{sec:methods}

Following a similar setup to \cite{hammer17}, we use the two-fluid version of the hydrodynamical code FARGO \citep{FARGO, zhu12} to simulate vortex formation induced by rapidly and slowly accreting planets (see Section~\ref{ssec:hydro}). We then produce synthetic images of our simulated asymmetric dust discs that can be compared to existing and future observations (see Section~\ref{ssec:rt-methods}).

% #### TWO-FLUID SIMULATIONS #### %

\subsection{Two-Fluid Hydrodynamics} \label{ssec:hydro}

\subsubsection{Numerical Methods} \label{sssec:hydro-overview}

FARGO uses the finite difference method and adaptive timesteps to solve the Navier-Stokes equations in 2-D cylindrical polar coordinates ($r$, $\phi$). The code takes advantage of the FARGO algorithm to speed up simulations by subtracting out the average azimuthal velocity when calculating the Courant-Friedrich-Levy condition that is used to determine the largest allowed timestep.

This variation of FARGO \citep{zhu12} builds upon the original code by adding a dust component as a secondary pressureless fluid. In addition to the source terms felt by the gas, the dust also feels radial and azimuthal drag forces due to its coupling with the gas. These terms are given by
\begin{equation} \label{eqn:rad_drift}
\frac{\partial v_\mathrm{r, d}}{\partial t} = - \frac{v_\mathrm{r, d} - v_\mathrm{r, g}}{t_\mathrm{s}},
\end{equation}
\begin{equation} \label{eqn:az_drift}
\frac{\partial v_\mathrm{\theta, d}}{\partial t} = - \frac{v_\mathrm{\theta, d} - v_\mathrm{\theta, g}}{t_\mathrm{s}},
\end{equation}
where $v_\mathrm{r}$ and $v_\mathrm{\theta}$ are the radial and azimuthal velocity components and $v_\mathrm{d}$ and $v_\mathrm{g}$ are the dust and gas components respectively. Since we only simulate small particles, the stopping time always falls in the Epstein regime \citep{weidenschilling77} and is given in the midplane by
\begin{equation} \label{eqn:stopping}
t_\mathrm{s} = \frac{\mathrm{St}}{\Omega} = \left( \frac{\pi}{2} \frac{\rho_\mathrm{d} s}{\Sigma_\mathrm{g}} \right) \frac{1}{\Omega},
\end{equation}
where $\rho_\mathrm{d}$ is the physical density of each dust grain (which we take to be 1 g / cm$^3$), $s$ is the size of each grain, $\Sigma_\mathrm{g}$ is the local surface density of the gas, and $\Omega$ is the local angular velocity. The Stokes number St is the dimensionless form of the stopping time. Besides being affected by drift, the dust can also diffuse through the disc due to turbulence. This effect acts as an extra source term \citep{clarke88} for the evolution of the density and is given by
\begin{equation} \label{eqn:diffusion}
\frac{\partial \Sigma_\mathrm{d}}{\partial t} = \nabla \cdot \left( D \Sigma_\mathrm{g} \nabla\left( \frac{\Sigma_\mathrm{d}}{\Sigma_\mathrm{g}}\right) \right),
\end{equation}
where the diffusion coefficient, $D$, can be approximated as the turbulent {\cbf viscosity}, $\nu$, for small particles with St $\ll 1$ \citep{youdin07}. The evolution of the gas is not affected by the dust since we do not include dust feedback.\footnote{However, the gas simulations are still not identical for each dust size because numerical noise propagates over time.} {\cbf We address the limitations of neglecting feedback in Sections~\ref{ssec:dust-results}~and~\ref{ssec:confirmation}.} For more details on the implementation of the dust component, see Section 2.1.2 of \cite{zhu12}.

%%Outline: \\
%%(1) We used two-fluid FARGO, which works like this. \\
%%(2) It is different because it has a dust component, which works like this. \\
%%(3) Planet Growth + Accretion onto Planet --- briefly \\
%%(4) Boundary Conditions \\

\subsubsection{Gas Component} \label{sssec:hydro-gas}

Each simulation begins with a locally isothermal gas disc having a radial density profile of $\Sigma_\mathrm{g}(r) = \Sigma_\mathrm{g,0} (r/r_\mathrm{p})^{-1}$, where $\Sigma_\mathrm{0}$ is the initial surface density at the orbital radius of the planet, $r_\mathrm{p}$. The value of $\Sigma_\mathrm{g,0}$ is set by the total disc mass in the domain, which we choose to be $M_\mathrm{d} = 0.002 = 2 M_\mathrm{p}$. We set the disc's aspect ratio to a constant value $h \equiv H/r = 0.06$,  which empirically maximizes the strength of the vortex for fixed planet and disc parameters \citep{fu14a}. A constant aspect ratio corresponds to a temperature profile that scales as $T(r) \propto r^{-1}$. The viscosity is set to $\nu = 10^{-7}$ in dimensionless units of $r_\mathrm{p}^2 \Omega_\mathrm{p}$. This value corresponds to $\alpha = 3 \times 10^{-5}$ near the location of the planet for the standard alpha prescription of $\nu = \alpha H^2 \Omega_\mathrm{p}$ \citep{alpha}. Our previous work found that this viscosity permitted slowly-growing planets to induce long-lived elongated vortices for the parameters used in this study \citep{hammer17}.

We simulate an annulus of a disc across a domain that spans $r \in [0.2, 5.7]r_\mathrm{p}$ in radius and $\phi = [0, 2 \pi]$ in azimuth. {\cbf The annulus is resolved by $N_\mathrm{r} \times N_\mathrm{\phi} = 2048 \times 3072$ arithmetically-spaced grid cells in the radial and azimuthal directions respectively, which is comparable to the grid sizes used in recent studies of dust evolution in vortices \citep[e.g.][]{miranda17, barge17, surville18}. This is high enough to resolve both the disc's scale height $H(r)$ and the planet's Hill radius by at least 22 cells in the vicinity of both the planet and the vortex. Additionally, the strong dependence of the vortex properties on the viscosity in \citep{hammer17} verifies that the resolution is sufficient for the prescribed viscosity to dominate over any numerical viscosity from the code.}

\subsubsection{Dust Component} \label{sssec:hydro-dust}

Alongside the gas disc, we initialize a dust fluid component that follows the same power law density profile over the same domain, except with a lower surface density of $\Sigma_\mathrm{d, 0} = \Sigma_\mathrm{g,0} / 100$. We do not prescribe a viscosity to the dust fluid, leaving it to diffuse only through the gas turbulence as described in Section~\ref{sssec:hydro-overview}. In a given simulation, the dust fluid represents particles of a single fixed size  specified as an input parameter (see Section~\ref{sssec:hydro-suite}).

\subsubsection{Planet} \label{sssec:hydro-planet}

At a radius of $r_p$, we place a Jupiter-mass planet ($M_\mathrm{p} = M_\mathrm{J}$) on a fixed circular orbit around a solar mass star ($M_\mathrm{*} = M_\mathrm{\odot}$) with an orbital angular velocity of $\Omega_\mathrm{p} = \sqrt{GM_\mathrm{\odot} / r_\mathrm{p}^3}$, where $G$ is the gravitational constant. This is equivalent to an orbital period of $T_p \equiv 2\pi / \Omega_\mathrm{p}$. For simplicity, we set $r_\mathrm{p} = \Omega_\mathrm{p} = G = M_\mathrm{*} = 1$ when used in the code. The planet's gravitational potential is given by 
\begin{equation} \label{eqn:potential}
\phi_\mathrm{p}(\mathbf{r}) = \frac{GM_\mathrm{p}}{\sqrt{(\mathbf{r} - {\mathbf{r}_\mathrm{\mathbf{p}}})^2 + r_\mathrm{s}^2}}
\end{equation}
where $r_\mathrm{s} = 0.6h$ is the smoothing length.

We prescribe the planet's mass, $m_\mathrm{p}$, as a function of time, $t$, to be: 
\begin{equation} \label{eqn:growth}
m_p(t) = M_p\times \begin{cases}
\sin^2{\left(\pi t / 2T_\mathrm{growth}\right)}  & t\leq T_\mathrm{growth}, \\
1 & t >   T_\mathrm{growth},
\end{cases}
\end{equation}
where $M_p=m_p(T_\mathrm{growth})$ is the planet's final mass and $T_\mathrm{growth}$ is the planet's growth timescale. We do not allow gas to accrete onto the planet. To avoid numerical issues that arise when mass accumulates within the planet's Hill sphere, we slowly remove this material following the prescription used by \cite{kley99}. The total mass removed over an entire simulation is always less than $5\%$ of the planet's final mass.

\subsubsection{Boundary Conditions} \label{sssec:hydro-boundaries}

At the radial boundaries of the disc, we utilize wave killing zones \citep[e.g.][]{deValBorro06} for $r~\in~[1, 1.25]r_\mathrm{in}$ and $r~\in~[0.84, 1]r_\mathrm{out}$ to damp the density and velocity fields in these regions towards the initial conditions for both the gas and the dust. We rapidly damp the outer zone on a timescale of 1/500th of the outermost orbital period so that this region is mostly unperturbed. The inner region is damped at a slower rate of 1/3rd of the innermost orbital period. We use periodic boundary conditions in the azimuthal direction.

\subsubsection{Simulation Suite} \label{sssec:hydro-suite}

We conduct {\cbf a total of eight simulations in our primary study, encompassing planets with two different growth timescales ($T_\mathrm{growth} = 10$ and $750~T_p$) and dust particles with four different sizes ($s =$ 0.3 mm, 1 mm, 3 mm, 1 cm)}. The largest grain size corresponds to an initial Stokes number, $\mathrm{St_0}$, of 0.076 at the location of the planet. For each particle size, we explore the differences in the contrast and azimuthal profiles of the dust vortices that form in each case, with a focus on the distinctive characteristics of the corresponding synthetic images at typical ALMA beam sizes.

%%We discuss how to combine the simulations of dust particles with different sizes into a single image in the next section.

%%Outline: \\
%%(1) Gas Disk setup \\
%%(2) Dust Disk setup \\
%%(3) Planet setup \\
%%(4) Set things equal to one 

% #### RADIATIVE TRANSFER SIMULATIONS #### %

\subsection{Synthetic Images} \label{ssec:rt-methods}

We generate synthetic intensity images from our simulations at {\cbf a cadence of one output per orbit} by calculating a map of the flux contributions from the dust surface density in each grid cell over a range of grain sizes, {\cbf a similar method to that used in other studies \citep[e.g.][]{bae16}}. Combining dust density fields is non-trivial since the gas vortices do not perfectly overlap between different simulations. {\cbf These deviations result from the different adaptive timesteps used by FARGO for different grain sizes. To align the density fields for different grains, we rotate them so that each gas vortex aligns with the one from the simulation of the intermediate 3 mm grains. We find the proper alignment angle by testing all integer angles between $0$ and $360^{\circ}$ and choosing the one with the lowest absolute difference in mass when subtracting each density field from the reference gas vortex at the same timestep. We checked the alignment manually and found this method is typically precise to within 5 degrees, often less. This level of precision is sufficient, given the relatively large beam sizes in the convolution step. With this alignment procedure, we can then combine the dust density grids corresponding to different grain sizes.}

{\cbf We combine the density maps from the four sizes that we simulated into a single composite density map with a number density distribution of $n = n_0(s/s_\mathrm{0})^{-p}$ where $p = 3.0$ and $n_0$ normalizes the initial gas-to-dust ratio to 100. Such a high power law index assumes significant grain growth in the vortex \citep{birnstiel10}. The maximum grain size is assumed to be 1 cm. In accordance with that power law, the weights applied to the density maps are as follows: $45\%$ to 1 cm, $37\%$ to 3 mm, $12\%$ to 1 mm, and $5.3\%$ to 0.3 mm. A lower power law index of -3.5 would reduce the magnitude of the peak in the resulting images, as smaller grains exhibit flatter dust distributions.}

% Power Law: [ 0.001       0.01632051  0.03745175  0.11843283  0.37451748  0.45227744]

The flux $F$ at a frequency $\nu$ is computed in each cell as
\begin{equation} \label{eqn:flux}
F_\mathrm{\nu}(r, \phi) = I_\mathrm{\nu}(r, \phi) \frac{r \delta r \delta \phi}{d^2},
\end{equation}
where $\delta r$ and $\delta \phi$ are the cell widths and the intensity $I_\mathrm{\nu}$ is given by
\begin{equation} \label{eqn:intensity}
I_\mathrm{\nu}(r, \phi) = B_{\nu}(T)[1 - e^{-\tau_{\nu}}].
\end{equation} 
The optical depth $\tau_{\nu}$ is
\begin{equation} \label{eqn:tau}
\tau_{\nu} = \Sigma_\mathrm{dust}(r, \phi; s) \kappa_{\nu},
\end{equation} 
where $\kappa_{\nu}$ is the opacity of a particular-size grain when observed at a particular frequency. These opacities are computed from the Jena database\footnote{http://www.astro.uni-jena.de/Laboratory/Database/databases.html} using Mie theory \citep{bohren83}, assuming that  the grains are magnesium-iron silicates \citep{jaeger94, dorschner95}.

The images are calculated at $\lambda = 0.87$ mm, which corresponds to ALMA Band 7. The distance to the system is set to $d = 140$ pc. To incorporate the effects of finite resolution found in observations, we convolve the images with a range of beam diameters from $0.071^{\prime \prime}$ to $0.284^{\prime \prime}$, corresponding to scales of 10 to 40 AU in the disc. Since most of the transition discs with observed asymmetries have large cavities, we neglect the inner disc ($r < r_\mathrm{p}$) to better mimic existing observations and to prevent the inner region from contaminating the outer disc when convolving with large beam sizes.

We scale our simulations by placing the planet at $r_\mathrm{p} =$ 20 AU. We also use a temperature profile of $T = T_0 (r / r_\mathrm{p})^{-1}$ consistent with our hydrodynamic simulations, where $T_0$ is set by the scale height of the disc and assumes a mean molecular weight of $\mu = 2.34$. The temperature of the star is set to $T_{\odot} = 5770$ K. The disc is viewed face-on with an inclination of $i = 0^{\circ}$.

{\cbf In our analysis of the synthetic images, we measure the offset of the peak from the center of the vortex directly from the image assuming no knowledge of the underlying gas vortex. We identify the edges of the vortex by choosing a fixed threshold relative to the intensity peak. From there, we assume the geometric center is at the midpoint of the edges.}

%%$ $\\
%%Outline: \\
%%(1) We used Paola's radiative transfer code to simulate observations. \\
%%(1a) Paola will explain this part. \\
%%(2) How to interpolate dust particle sizes \\
%%(3) Other parameters used in RT simulations \\

% #### RESULTS #### %

\section{Results} \label{sec:results}

\subsection{Overview} \label{ssec:overview}

%%Figure 1: \\
%%Gas and Dust Density Plots (3 different grain sizes) \\
%%(Two 4-panel plots --- either similar to time evolution plots from the last paper, or 2x2) \\
%%
%%Figure 2: \\
%%Azimuthal Dust Density Profiles at a specific orbit (w/ Lyra + Lin for Comparison) \\
%%
%%Figure 3: \\
%%Azimuthal Extents as a function of time (for different density contours) \\
%%(or maybe just figure out a better way to highlight the azimuthal extents as a function of density in the previous figure? -- such as coloring each line segment differently depending on which density contour it is above) \\
%%
%%Figure 4: \\
%%Radiative Transfer Images (convolved only) \\
%%
%%Figure 5: \\
%%Collage of Vortex Candidates \\
%%
%%Outline: \\
%%(1) Our simulations show that [bulletpoint comparisons of the elongated vortices to the concentrated vortices] \\
%%(1a) Contrast \\
%%(1b) Azimuthal Extents \\
%%(1c) Gaussian power fitting to radial and azimuthal profiles \\
%%
%%
%%
%%--------- As a secondary thing, does the radial gaussian have the same power as the azimuthal gaussian (says something about dominance of azimuthal drift versus radial drift) \\

% #### RESULTS #### %

\begin{figure*} 
\centering
\includegraphics[width=0.98\textwidth]{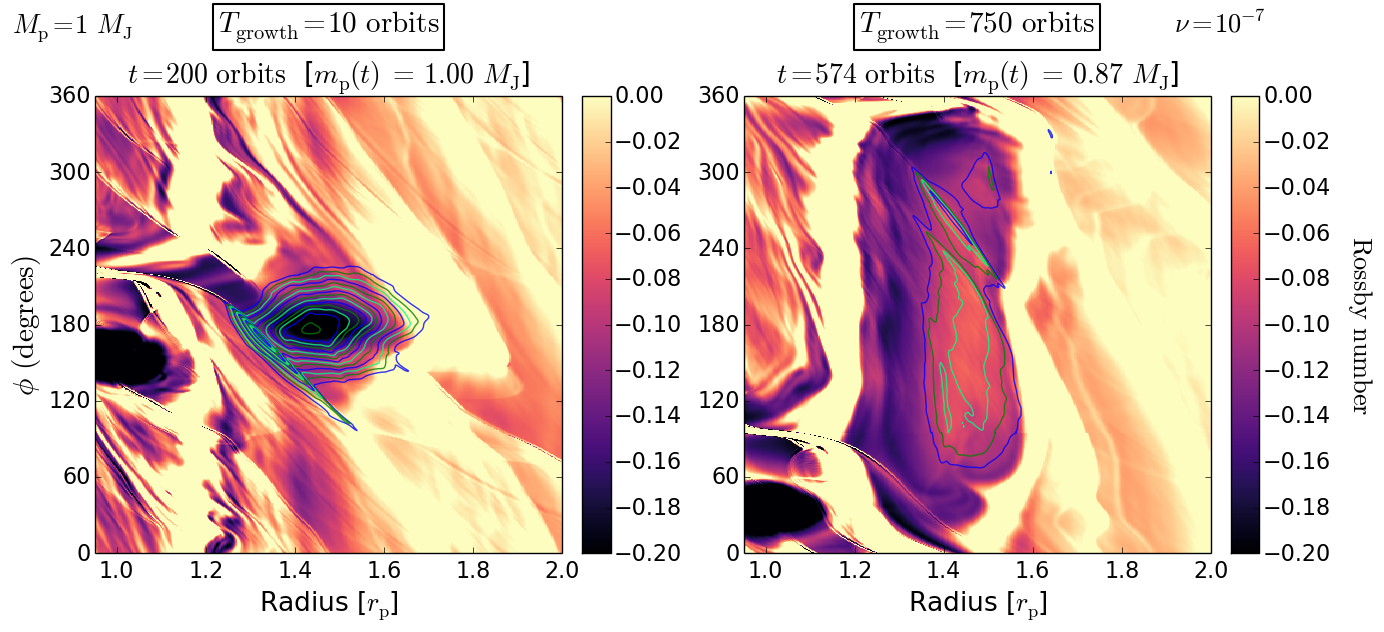} 
\caption{Rossby number snapshots of the concentrated vortex ($T_\mathrm{growth} = 10$ planet orbits; left panel) and the elongated vortex ($T_\mathrm{growth} = 750$ planet orbits; right panel). Gas density contours are overlaid on each plot corresponding to $\Sigma_\mathrm{g} / \Sigma_\mathrm{g,0}$ = 1.0, 1.1, 1.2, ..., 2.3. The concentrated vortex has a vorticity minimum at the center and a smooth pressure bump, both of which facilitate dust trapping at the center. The elongated vortex lacks these features, leaving the dust it traps much more spread out in the azimuthal direction. \textit{The Rossby number is defined as $\mathrm{Ro} \equiv [\nabla \times (\mathbf{v} - {\mathbf{v}_\mathrm{\mathbf{K}}})]_\mathrm{z} / 2 \Omega$, where $\mathbf{v}$ is the gas velocity and $\mathbf{v}_\mathrm{\mathbf{K}}$ is the Keplerian gas velocity.}}
\label{fig:vorticity}
\end{figure*}

\begin{figure*} 
\centering
\includegraphics[width=0.95\textwidth]{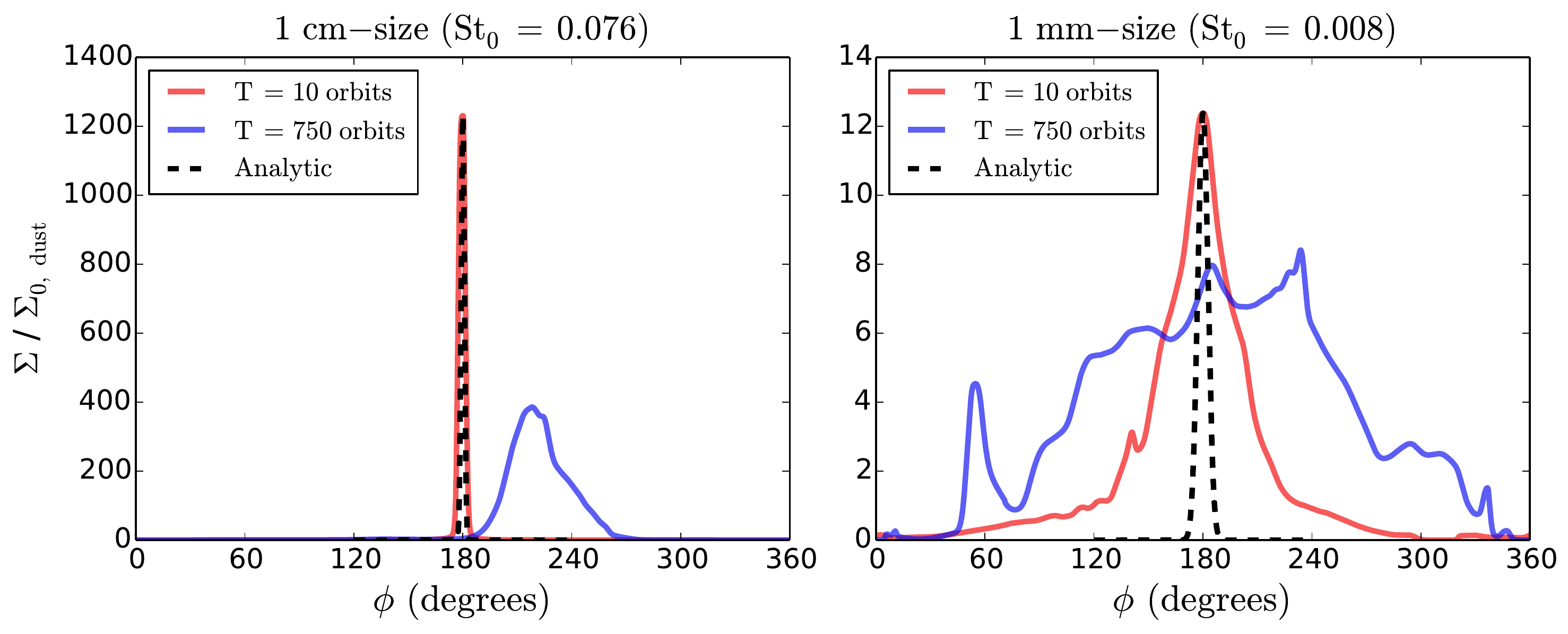}
\caption{Central azimuthal dust density profiles of the concentrated vortex ($T_\mathrm{growth} = 10$ planet orbits; shown in red) and the elongated vortex ($T_\mathrm{growth}~=~750$ planet orbits; shown in blue) for cm-sized (left panel) and mm-sized dust (right panel). Analytic models (black dashed lines; see Section~\ref{sssec:dust-model}) through the center are shown for comparison. For both grain sizes, the concentrated vortices trap dust in Gaussian profiles in the vortex center. In contrast, the elongated vortices have off-center peaks and much wider profiles. \textit{Snapshots are the same as in Figure~\ref{fig:vorticity}. }}
\label{fig:first_comparison}
\end{figure*}

{\cbf
We find that elongated vortices do not trap dust in the same manner as concentrated vortices, making them easy to distinguish in synthetic images with sufficient resolution.
It is widely accepted that concentrated vortices trap dust in pressure maxima at their centers \citep{barge95, tanga96, johansen04}, provided that they satisfy the condition $|\delta{\bf u}| \geq c_\mathrm{s}^2 / 2v_\mathrm{K}$, where $\delta{\bf u}$ is the velocity perturbation, $v_\mathrm{K}$ is the Keplerian velocity, and $c_\mathrm{s}$ is the sound speed. These pressure maxima also correspond to vorticity minima. Additionally, this relation assumes that a vortex maintains geostrophic balance between the pressure gradient and Coriolis force \citep[see review by][and references therein]{lovelace14}. 

On the contrary, elongated vortices do not have the same type of uniform structure (see Figure~\ref{fig:vorticity}). While they maintain a sharp pressure gradient towards their boundaries, their interior pressure profiles are mostly flat. The azimuthal gradient is negligible except near the edges, while the radial gradient also flattens out near the center over a wider radial range than in the concentrated case. Without steeper pressure gradients, no vorticity minimum arises at the center. Moreover, the interior profiles are ever-changing due to the repeated interactions between the vortex and the planet's spiral density waves. These waves act as a larger perturbation $\delta \Sigma / \Sigma_\mathrm{g}$ through the vortex in the elongated case. They also have a more pronounced effect of altering the overall structure of the gas in the elongated case compared to the concentrated case, which we suspect is due to shocks. As a result of its flatter and time-variable structure in the gas, an elongated vortex traps dust away from its center, unlike a more idealized concentrated vortex.

As usual, the dynamics of dust particles in an elongated vortex are strongly dependent on particle size. Instead of concentrating at the center, dust of all sizes circulates around the vortex on timescales of a few dozen orbits. Larger dust particles concentrate more and circulate closer to the vortex center. Only the largest particles ($s \geq$ 3 mm) exhibit pronounced peaks, while smaller dust particles ($s \leq$ 1 mm) are spread out across nearly the entire vortex over a width of $\geq 180^{\circ}$. Even though larger dust particles are still the most concentrated in elongated vortices, they are also much more spread out compared to dust of the same size in concentrated vortices (see Figure 2). Furthermore, dust preferentially circulates towards the front of the vortex in the direction of the planet's orbit, possibly due to the motion of the spiral density waves. The circulation of the dust as well as the interactions with the spiral waves make the azimuthal profiles of elongated vortices highly variable over time.

The circulation of the dust across an elongated vortex produces a variety of signatures that distinguish it from a concentrated vortex. These include
\begin{enumerate}[leftmargin=1cm, itemindent=-0.2cm]
  \item wide azimuthal extents
  \item off-center peaks, and
  \item a skewness (lack of symmetry) about the peak, as well as
  \item double peaks in newly-formed vortices.
\end{enumerate}
Observations with a beam size equal to or less than the semimajor axis of a putative planet -- that is, a beam size much smaller than the radial location of the vortex itself -- are needed to identify these features. With larger beam sizes, the edge of the vortex begins to blend with the background. As a result, it is difficult to find the azimuthal edges of the vortex, which in turn prevents precise measurements of the azimuthal extent as well as the magnitude of the peak offset, the latter of which is easily underestimated at lower resolution.
}

\subsection{Dust Vortices} \label{ssec:dust-results}

\begin{figure} 
\centering
\includegraphics[width=0.48\textwidth]{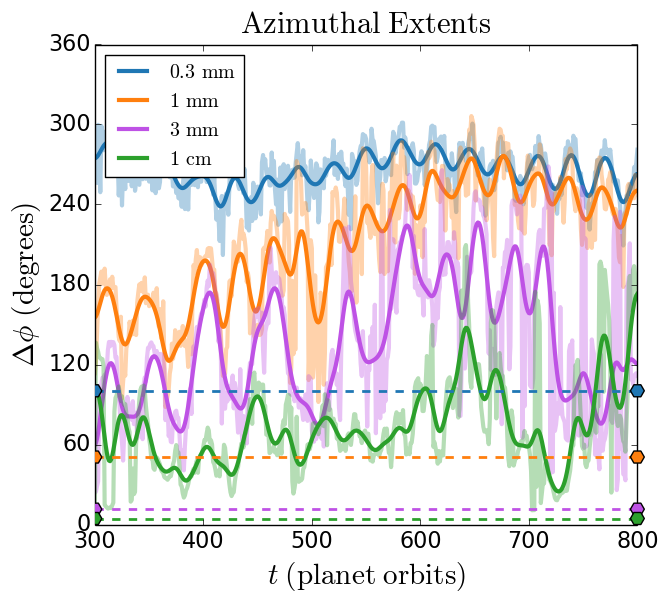}
\caption{Extents of the elongated vortex over time for four different grain sizes. The extents of the concentrated vortex at a single time of $t = 200$ are denoted with dashed lines for comparison.
}
\label{fig:extents}
\end{figure}

\begin{figure*} 
\centering
\includegraphics[width=0.48\textwidth]{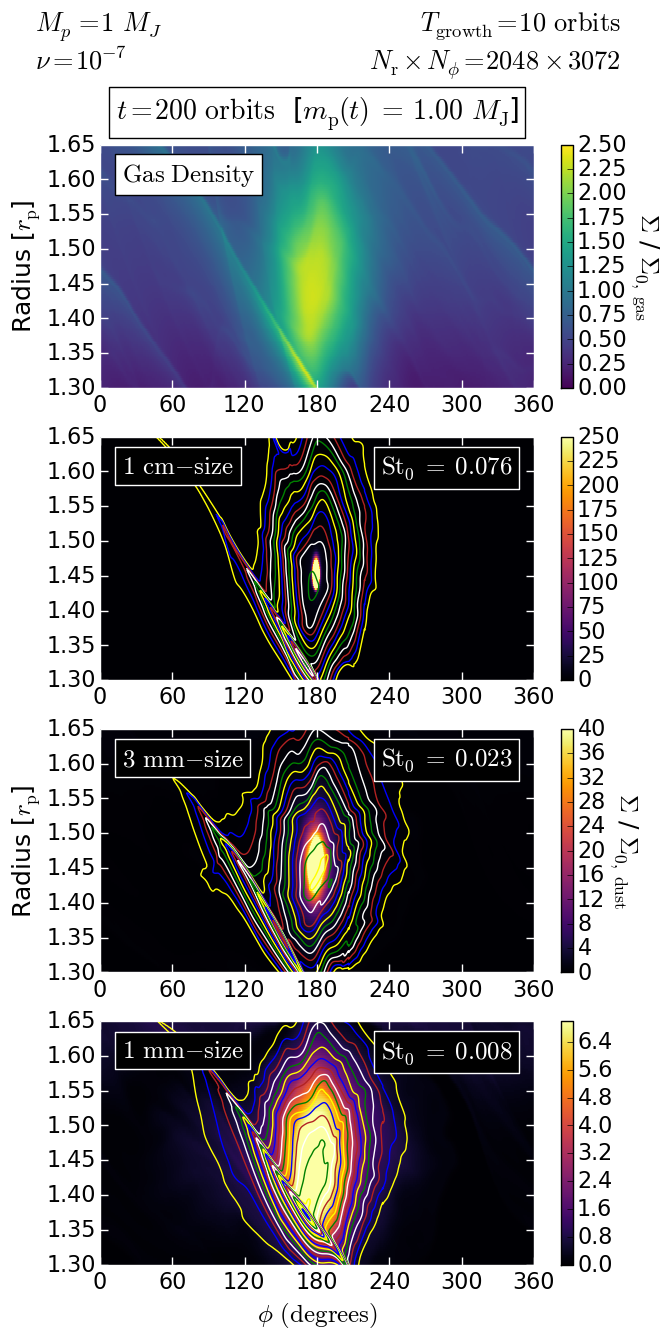}
\hspace*{1.0em}
\includegraphics[width=0.48\textwidth]{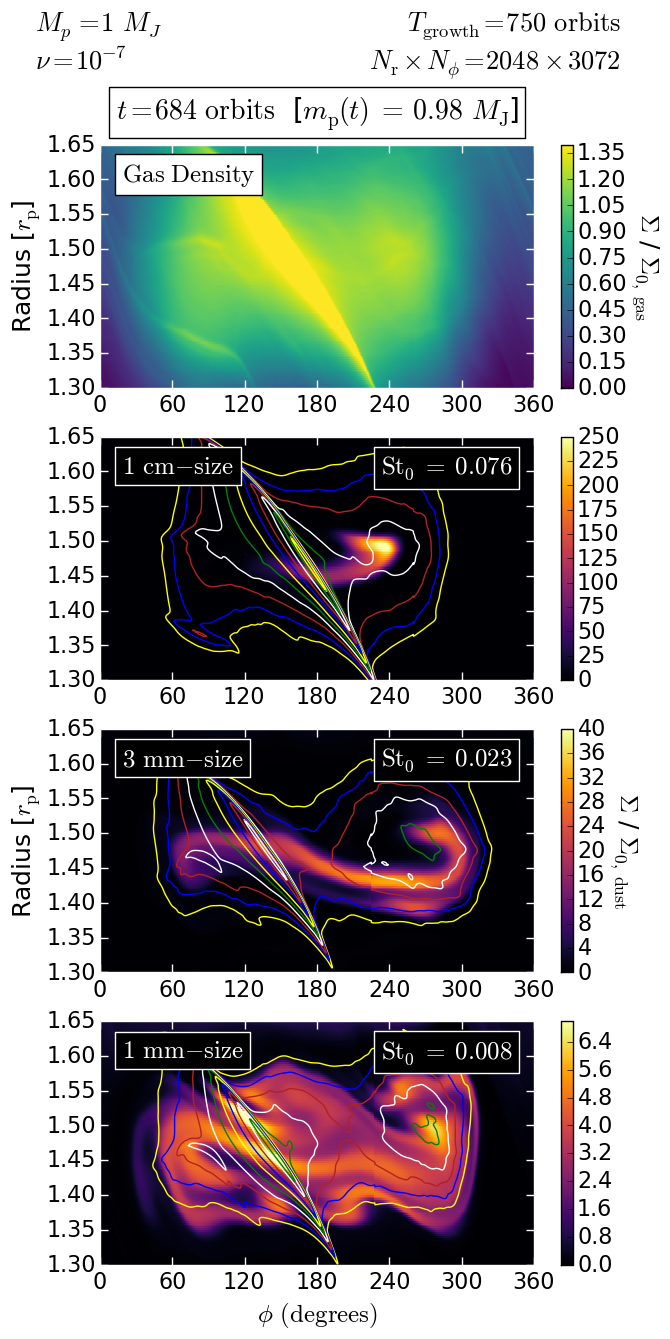}
%\includegraphics[width=0.8\textwidth]{figures/densityMap_0300.png}
%\vspace*{2.5em} \\
%\hspace*{2.5em}
%\includegraphics[width=0.85\textwidth]{figures/azimuthalDensityProfiles_0300.pdf}
\caption{Density maps of the concentrated vortex ($T_\mathrm{growth} = 10$ planet orbits; left panels) and the elongated vortex ($T_\mathrm{growth} = 750$ planet orbits; right panels). The top frame shows the gas distribution; the bottom frames show the dust distributions for the three largest grain sizes in our study: 1 cm, 3 mm, and 1 mm. Gas density contours are overlaid on each dust density map corresponding to $\Sigma_\mathrm{g} / \Sigma_\mathrm{g,0}$ = 0.9, 1.0, 1.1, ..., 2.3. In the elongated vortex, all three dust maps show dust circulating back towards the center from the front of the vortex. Larger grains are more concentrated in the peak and have narrower extents. \textit{Movies are also included.}}
\label{fig:triple_density_maps}
\end{figure*}

{\cbf
As the dust circulates around an elongated vortex, the azimuthal extents of different-sized grains do not stay fixed over time (see Figure~\ref{fig:extents}). This behavior is very different from that of concentrated vortices, where the extents hardly change. Furthermore, larger grains in elongated vortices are not necessarily much more concentrated than smaller grains. The extents of a concentrated vortex are roughly $60^{\circ}$ for mm grains, and less than $10^{\circ}$ for cm grains. There are also clear gaps between any two successive grain sizes from our simulations. In contrast, the extents of an elongated vortex vary between $60^{\circ}$ to $120^{\circ}$ for cm grains, and are between $90^{\circ}$ to $210^{\circ}$ for 3 mm grains. While the extent of the 1 mm grains is initially much narrower than that of the 0.3 mm grains by more than $90^{\circ}$, they both converge to a little above $240^{\circ}$. There are also rare instances where the 3 mm grains are slightly more concentrated than the 1 cm grains, or slightly less concentrated than the 1 mm grains. These extents are measured with a threshold of $40\%$ of the peak; other choices do not change the qualitative results.
}

%%% ### Outline: New Paragraph 2 ### %%%
%% (2) circulation patterns, including overlap --- DONE \\

{\cbf
The circulation of the dust in an elongated vortex is also marginally size-dependent (see Figure~\ref{fig:triple_density_maps}). Each period of circulation is roughly a few to several dozen orbits; however, the dust density peak rarely completes a closed trajectory around the vortex. These open trajectories are due to the narrowness of the vortex's radial extent and the non-uniformity of the gas in the vortex interior, the latter of which is partially attributed to shocks from the planet's spiral density waves. These features make it difficult to quantify the discrepancies between sizes. Nonetheless, with larger grains collecting closer to the center in both the radial and azimuthal directions, the peaks of different-sized grains do not perfectly overlap. When the vortex forms, the peaks are initially aligned. However, as the gas structure evolves, small differences in the peak locations propagate over time. Nevertheless, the offset between peaks at different grain sizes does not continue to increase indefinitely, which we suspect is due to the spiral waves inhibiting the motion of the dust towards the tail of the vortex. The small azimuthal discrepancies between these peaks flatten out the intensity peak in the resulting synthetic images.
}

%%% ### Outline: New Paragraph 3 ### %%%
%% (3) different sized grains (peaks and shifts)

{\cbf
The telltale signature of dust circulation in an elongated vortex is the movement of the peak away from the center of the vortex. Only the dust density peaks for the largest grains ($s \geq 3$ mm) stand out. These peaks are shifted towards the front of the vortex more than half of the time. In particular, once the peak for the cm grains moves towards the front of the vortex, it can stay circulating at the center or ahead for several hundred orbits. In contrast, on the rare occasions when the peak for these grains moves towards the back of the vortex, it will circulate back to the center or the front within twenty orbits. We interpret this behavior as evidence that the spiral density waves are helping carry the dust towards the front of the vortex in general. Peak offsets more than~$30^{\circ}$ ahead of center are common, while the largest offsets can exceed $60^{\circ}$. In comparison to other causes of offsets, the observed shifts exceeding $30^{\circ}$ are larger than those that might arise due to self-gravity and the indirect force for all particles with $s < 5$ cm \citep{baruteau16}, as well as those caused by dust feedback \citep{barge17}. We cannot conclude whether incorporating dust feedback would enhance or diminish these effects, as that would depend on whether it induces azimuthal drift towards or away from the center of the vortex.
}

%%% ### Outline: New Paragraph 5 ### %%%
%% (5) feedback relevant

Even though the elongated vortex has less pronounced peaks than the concentrated vortex, it still has dust-to-gas ratios of order unity. In Figure~\ref{fig:triple_density_maps}, the composite density field of the concentrated vortex achieves $\Sigma_\mathrm{d} / \Sigma_\mathrm{g} = 5.8$, while the elongated vortex achieves a lower $\Sigma_\mathrm{d} / \Sigma_\mathrm{g} = 1.6$. On longer timescales, both vortices continue to accumulate dust unimpeded as the dust drifts inwards through the disk, although the magnitude of the peak in the elongated vortex increases at a much slower rate than in the concentrated vortex. With these high dust concentrations ($\Sigma_\mathrm{d} / \Sigma_\mathrm{g} > 1$), dust feedback may reduce the lifetimes of the vortices or further widen their azimuthal extents \citep{fu14b, crnkovic15}, effects which we leave to future studies to investigate.

\subsubsection{Analytic Dust-Trapping Model} \label{sssec:dust-model}

In \cite{hammer17}, we used the analytic dust trapping models of \cite{lyra13} to estimate dust profiles in both concentrated and elongated vortices. Perhaps unsurprisingly, our dust simulations show wild departures from the analytic models in the elongated case. This is consistent with their messier internal velocity and density structures, which do not feature the closed pressure contours and streamlines of Kida vortices \citep{kida81}. To illustrate the discrepancy, we nevertheless compare our simulations with the analytic models below. Kida vortices are characterized by a constant aspect ratio, which we estimate as
\begin{equation} \label{eqn:aspect} 
\chi = r \Delta \phi / \Delta r,
\end{equation}
where  $r\Delta \phi$ and $\Delta r$ are the azimuthal and radial extents of the gas vortex respectively, which we measure from our simulations.

We then use \citeauthor{lyra13}'s model for the dust distribution across a vortex given by
\begin{equation} \label{eqn:dust} 
\rho_\mathrm{dust}(a) = \rho_\mathrm{0} \exp
\bigg\{ -\frac{a^2 f^2(\chi)}{2H^2}(S + 1)\bigg\} 
\end{equation}
(see their Equation 64) to derive the expected azimuthal profile. Here, $a$ is the semi-minor axis of an elliptical density contour inside the vortex; $\rho_0$ is the peak dust density in the vortex at a particular orbit in the simulation; $S =$ St$/ \delta$ is the ratio of the particles' Stokes number to the dimensionless diffusion coefficient in the vortex; and $f(\chi)$ is a scale function of order unity (see their Equation 35). Unlike in \cite{hammer17}, we select $\delta =  3 \times 10^{-5}$ to be the local value of the $\alpha$-viscosity, taken from our simulations. We do not presume any additional turbulence due to the presence of the vortex since we did not add any extra turbulence in the simulation.

{\cbf In Figure~\ref{fig:first_comparison}, we compare the azimuthal density distributions at the radial center of the vortex for dust with $s=1$ cm and 1 mm to the analytic models of the concentrated vortex. (With the dependence on $\chi$ in Equation~\ref{eqn:dust}, the analytic profiles of the elongated vortex are even narrower than those of the concentrated vortex shown in Figure~\ref{fig:first_comparison}.)} We find that the concentrated vortices have Gaussian profiles that match up rather well with the analytic profiles, aside from being slightly wider than predicted. Interestingly, the analytic models for the two largest grain sizes would {\cbf very closely} agree with the simulations if they assumed {\cbf a factor of five} higher level of diffusion ($\delta =  1.5 \times 10^{-4}$) compared to the prescribed value used in the simulations.

On the other hand, the profiles in the elongated vortices disagree qualitatively and quantitatively. The profiles for the cm-sized grains have much wider peaks that lack the symmetry of a Gaussian, while all other smaller simulated grain sizes have much flatter profiles. As a result, we advocate caution in applying analytic models to non-ideal vortices.

%This difference is not surprising given that the elongated vortices also have much flatter azimuthal gas density profiles with more rectangular streamline patterns compared to the simple assumption of elliptical streamlines that this model presumes.

%%$ $ \\
%%Outline: \\
%%(1) Compare dust vortices to gas vortices. \\
%%(1a) They have the same shape, but the dust distributions are more concentrated and higher amplitudes due to the vortex being efficient at trapping dust. \\
%%(2) Compare dust distributions at different wavelengths \\
%%(2a) Reference dust density figures. \\
%%(3) Compare two different growth times. \\
%%------ For azimuthal profile, fit to Gaussian of arbitrary power (like van der Marel 2015) -- or is this for the intensity? check the paper again... \\
%%------ As a secondary thing, does the radial gaussian have the same power as the azimuthal gaussian (says something about dominance of azimuthal drift versus radial drift) \\

% #### RADIATIVE TRANSFER RESULTS #### %

\subsection{Synthetic Images} \label{ssec:si-results}

\begin{figure*} 
\centering
\includegraphics[width=0.90\textwidth]{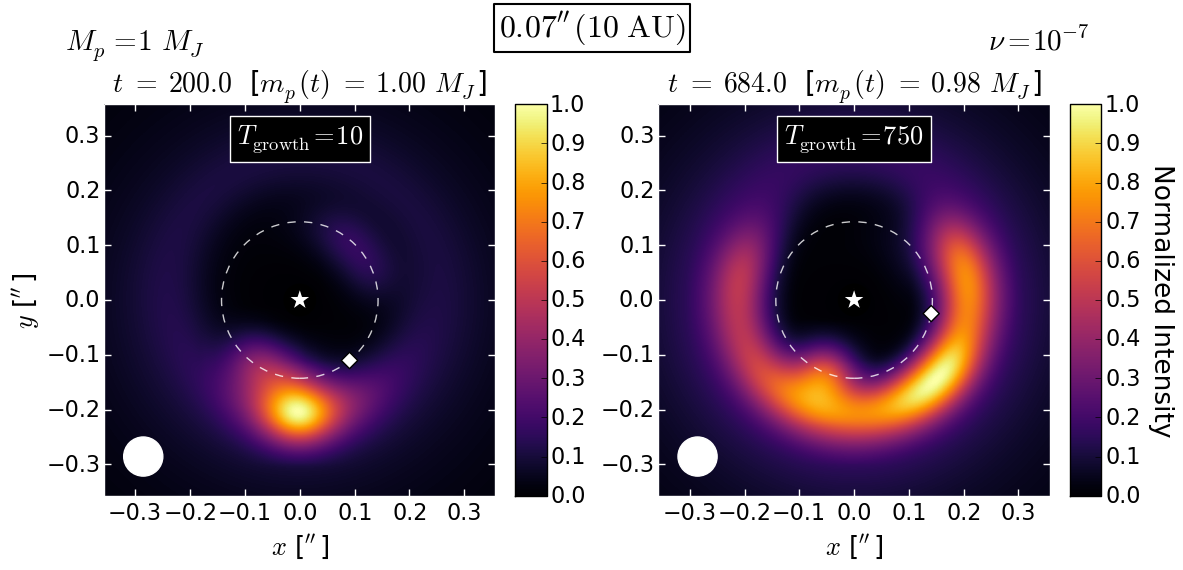} \\
%\vspace*{1.5em}
\includegraphics[width=0.47\textwidth]{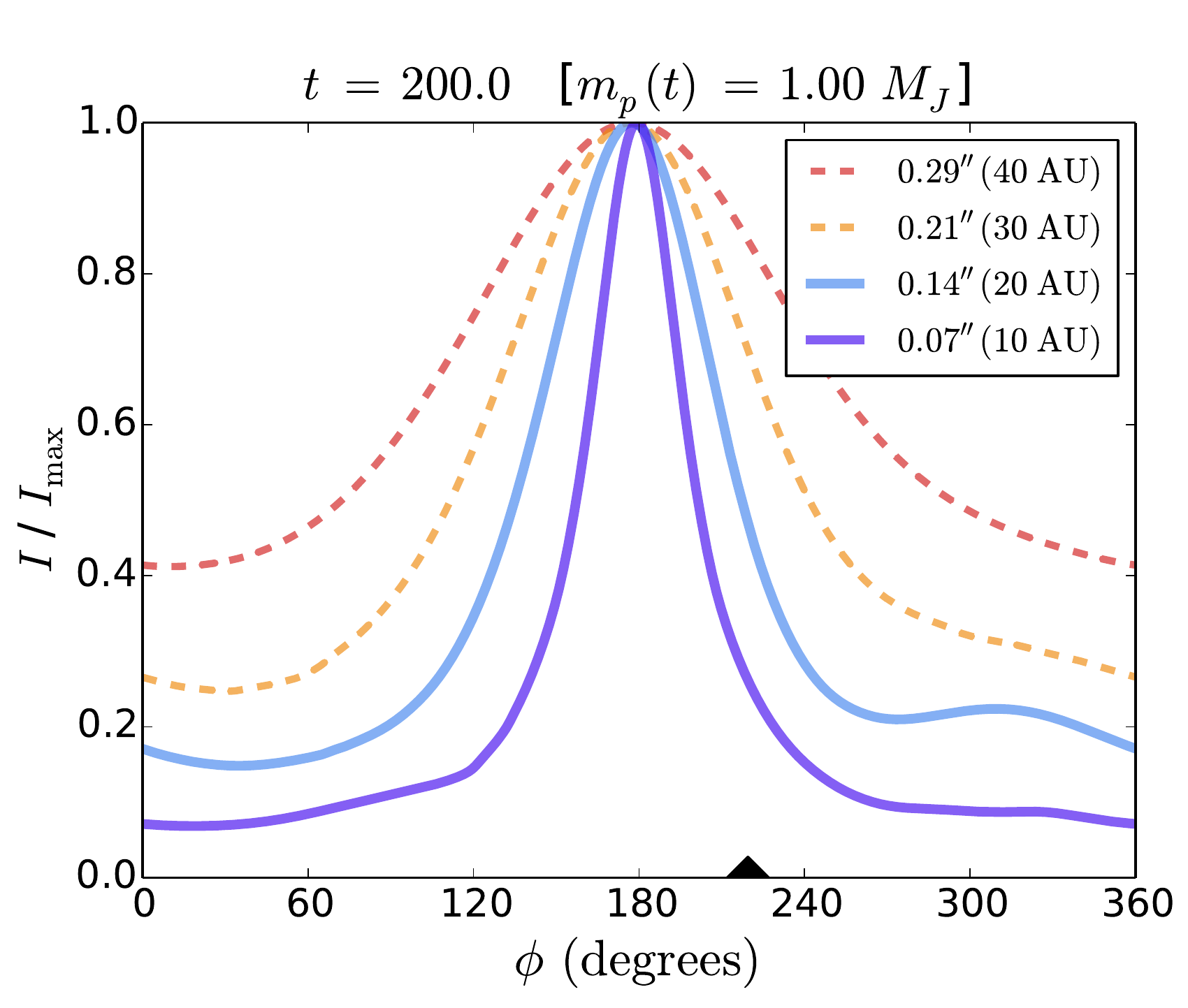}
%\hspace*{0.5em}
\includegraphics[width=0.47\textwidth]{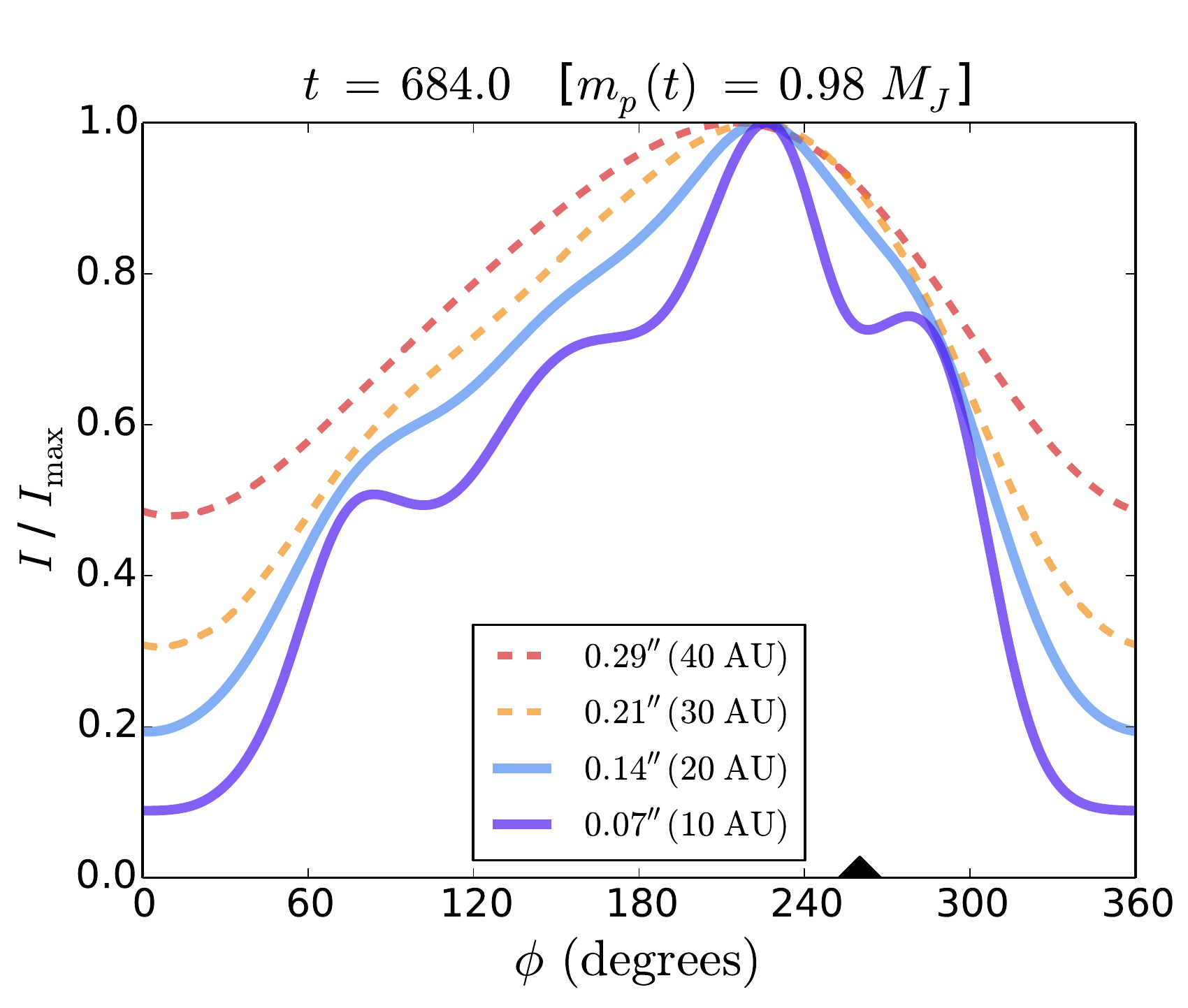}
\caption{Top: Synthetic images of the concentrated vortex ($T_\mathrm{growth} = 10$ planet orbits; left panels) and the elongated vortex ($T_\mathrm{growth} = 750$ planet orbits; right panels). Bottom: Comparison of central azimuthal intensity profiles for four different beam sizes. The black triangle indicates the location of the planet. With the smallest beam size, the peak in the elongated vortex is $40^{\circ}$ off-center. \textit{Movies of both types of plots of the elongated vortex are also included.}}
\label{fig:synthetic_images}
\end{figure*}

\begin{figure} 
\centering
\includegraphics[width=0.48\textwidth]{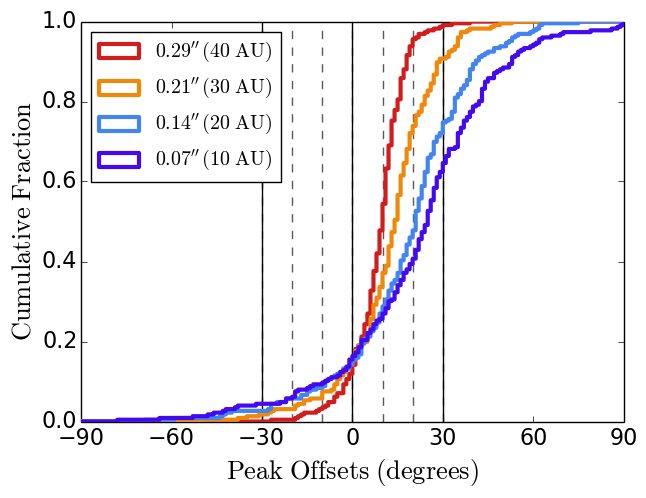}
\caption{Histogram of the peak offsets in elongated vortices over time for four different beam diameters. Larger offsets are much more frequent with higher resolution.
}
\label{fig:peak_histogram}
\end{figure}

%%Topics: \\
%%(0) intro \\
%%(1) distribution of peak offsets ---  find necessary beam size \\
%%(2) extents ---  find necessary beam size \\
%%(3) contributions to the peak from each grain size --- in both cases \\
%%(4) variability --- is current description wrong? (do high-cadence runs?) \\

%%% ### Outline: New Paragraph 0 ### %%%
%% (0) intro

{\cbf
The circulation of dust in an elongated vortex creates several signatures in our synthetic images, the most notable of which is the off-center peaks that are naturally accompanied by a skewness in the azimuthal profiles. The extent can also distinguish an elongated vortex from a concentrated one, provided that it is possible to identify the vortex edges. Identifying the largest peak offsets as well as measuring the wider azimuthal extents both typically require a beam size equal to or less than the planet's semimajor axis --- that is, roughly two-thirds the radial location of the peak in the asymmetry (see Figure~\ref{fig:synthetic_images}).
}

%%% ### Outline: New Paragraph 1 ### %%%
%% (1) distribution of peak offsets ---  find necessary beam size \\

{\cbf
The largest offsets of more than $30^{\circ}$ are regularly identifiable with this high of a resolution. Figure~\ref{fig:peak_histogram} shows the distribution of peak offsets for four different beam diameters: 10 AU, 20 AU, 30 AU, and 40 AU -- where the planet is located at 20 AU. With a beam diameter of the planet's separation (1.0 $r_\mathrm{p}$; equivalent to $0.14^{\prime \prime}$), shifts of more than $30^{\circ}$ in either direction occur $30\%$ of the time. This number drops to $10\%$ for a diameter of 30 AU (1.5 $r_\mathrm{p}$; equivalent to the separation of the vortex) and less than $2\%$ for twice the planet's semimajor axis. Relatively large shifts of $>20^{\circ}$ still appear in $60\%$ of snapshots for a beam size of the planet's separation, a number that drops to $20\%$ for a beam of the vortex's separation. However, we caution that the offsets in this more moderate range could also be attributed to the indirect force without self-gravity, an effect that yields maximum shifts of $25^{\circ}$ \citep{baruteau16}. Additionally, larger beam sizes can be insufficient for resolving secondary peaks (see Section~\ref{sssec:double-peaks}). For these reasons, we recommend a beam size of at most the planet's separation to measure the peak offset instead of the vortex's separation or larger.
}

%%% ### Outline: New Paragraph 2 ### %%%
%% (2) extents ---  find necessary beam size \\

{\cbf
High resolution is needed to measure a vortex's azimuthal extent as well. The bottom panels of Figure~\ref{fig:synthetic_images} shows the azimuthal profile across the radial center of a vortex for a single snapshot that is observed with a range of beam sizes from 10 to 40 AU, as with Figure~\ref{fig:peak_histogram}. With higher resolution, the sharpness of the edges makes it possible to identify their location with a good choice of threshold. Lower resolution at or above the separation of the vortex blend the edges with the background too much to pinpoint their location. This lack of precision makes it difficult to tell if the vortex has an elongated azimuthal extent, and is also the reason for the lower resolution underestimating the magnitude of the peak offsets. Moreover, the concentrated vortex also starts to appear elongated at these resolutions. We also caution that observations of disks that are not face-on may require even higher resolution to measure a vortex's azimuthal extent.
}

%%% ### Outline: New Paragraph 3 ### %%%
%% (3) contributions to the peak from each grain size --- in both cases \\

{\cbf
The appearance of the azimuthal profiles is dependent on the choice of grain size distribution. We can estimate the contributions of the grain sizes by calculating the peak intensity after separately removing each grain size from the images. Leaving out the 1 cm grains results in an average of a $29\%$ reduction in the peak of the elongated vortex, while removing the 3 mm grains leads to a $30\%$ reduction, the 1 mm grains a $17\%$ reduction, and the 0.3 mm grains a $23\%$ reduction. With their sharper distributions and higher contribution levels, the 1 cm and 3 mm grains are therefore most responsible for the properties of the peak. The azimuthal profiles of a vortex with a flatter grain size distribution such as $p = 3.5$ would be more dominated by smaller grains near $s \approx 3$ mm. Furthermore, we note that the relatively jagged azimuthal intensity profile shown in the right panel of Figure~\ref{fig:synthetic_images} for the smallest beam size might appear smoother had we simulated a more continuous distribution of grain sizes. Additionally, a continuous size distribution might also lead to vortices frequently having flatter intensity peaks due to the slight differences in the shifts of different-sized grains.
}

%% [0 24.81191697  25.96637237  15.00764272  20.5736164]

%%% ### Outline: New Paragraph 4 ### %%%
%% (4) double-peaked structures \\

% Moved to separate sub-section in this section.

%%% ### Outline: New Paragraph 5 ### %%%
%% (5) variability --- is current description wrong? (do high-cadence runs) \\

{\cbf
The circulation of dust in an elongated vortex can lead to quick changes in the azimuthal profiles of a vortex. Even though the circulation period itself is very slow, factors such as interactions with the planet's spiral density waves, the spreading of the peak, and dust motion near the azimuthal edges can cause significant variability within a single orbit. As a result of these effects, the location of the peak relative to the vortex center changes by more than $15^{\circ}$ in $18\%$ of snapshots taken at successive planet orbits with a beam diameter of 10 AU (0.5 $r_\mathrm{p}$; equivalent to $0.07^{\prime \prime}$). It may be possible to observe these changes in vortices induced by planets in the inner disk on timescales on the order of decades. Nonetheless, a lack of variability is still more common, with $42\%$ of successive snapshots having the peak unchanged to within $5^{\circ}$.
}

\subsubsection{Double-peaked Structures} \label{sssec:double-peaks}

%%% ### Outline: New Paragraph 1 ### %%%
%% (1) double peaked structures

\begin{figure*} 
\centering
\includegraphics[width=0.48\textwidth]{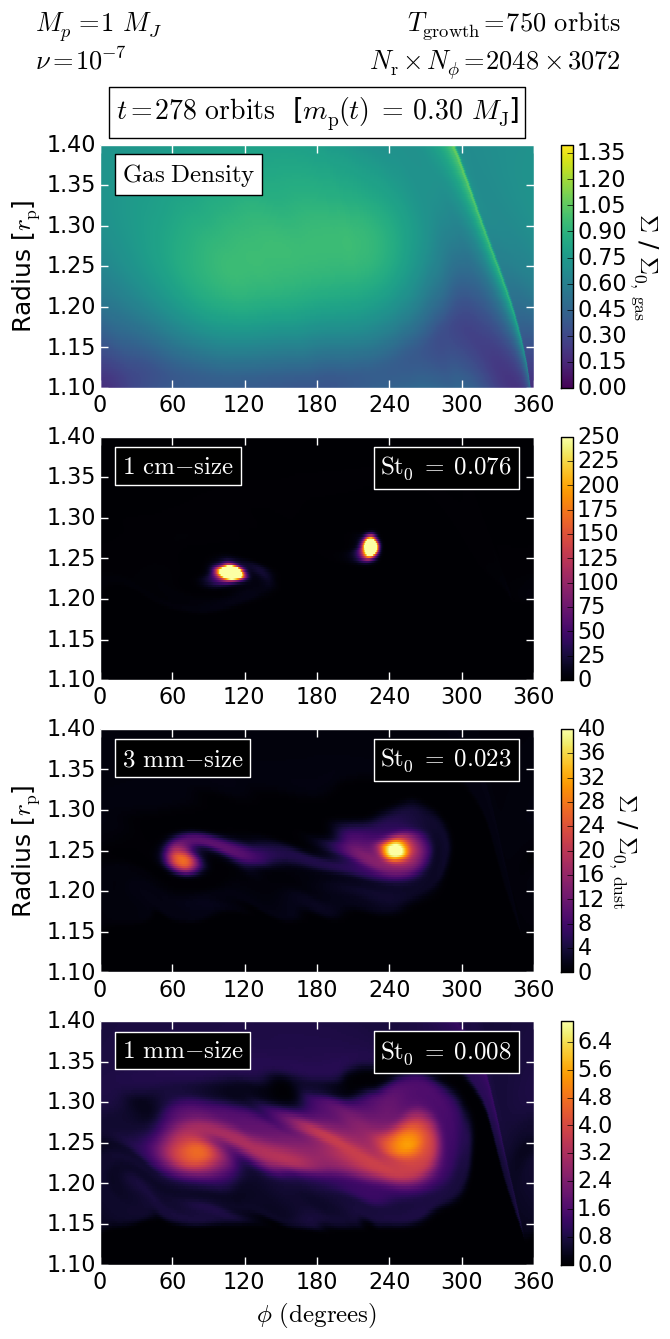}
\hspace*{1.0em}
\includegraphics[width=0.48\textwidth]{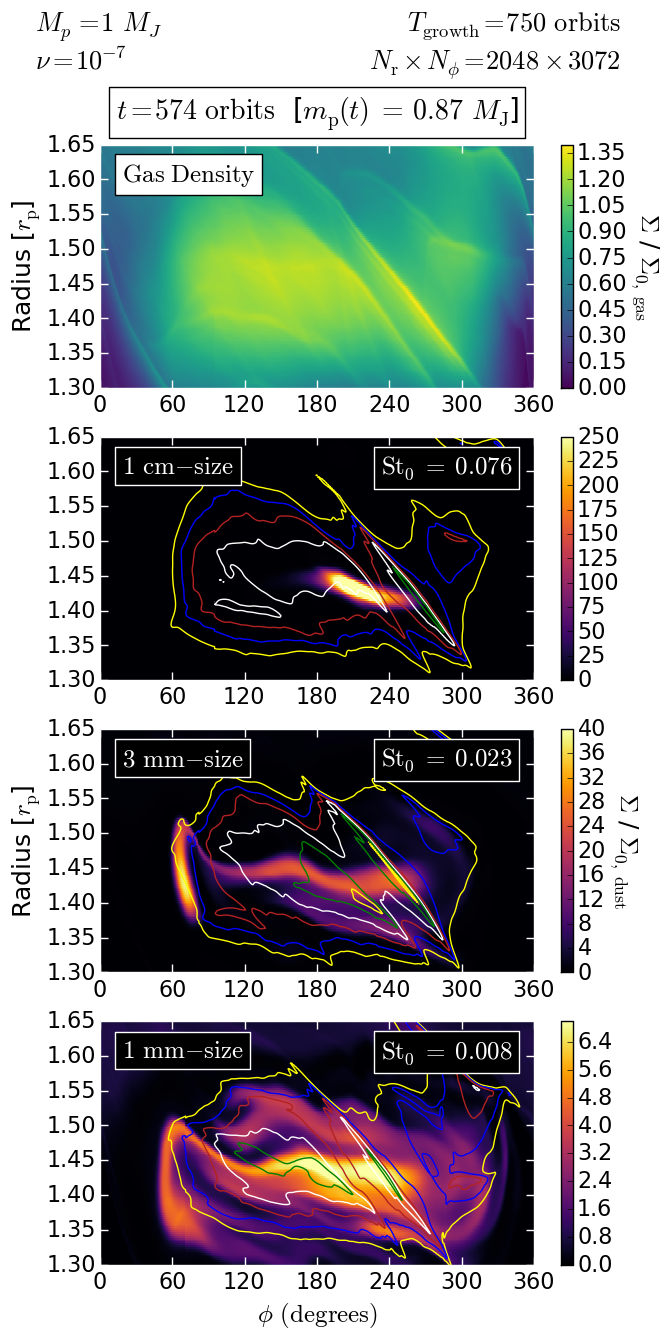}
%\includegraphics[width=0.8\textwidth]{figures/densityMap_0300.png}
%\vspace*{2.5em} \\
%\hspace*{2.5em}
%\includegraphics[width=0.85\textwidth]{figures/azimuthalDensityProfiles_0300.pdf}
\caption{Density maps of the elongated vortex ($T_\mathrm{growth} = 750$ planet orbits) that exhibit double peaks in the corresponding synthetic images. The top frame shows the gas distribution; the bottom frames show the dust distributions for the three largest grain sizes in our study: 1 cm, 3 mm, and 1 mm. The left panels show an earlier snapshot ($t = 278$; $m_\mathrm{p}(t) = 0.30\ M_\mathrm{Jup}$) within 50 orbits of the initial vortices merging into an $m=1$ mode. The right panels show a later snapshot ($t = 574$; $m_\mathrm{p}(t) = 0.87\ M_\mathrm{Jup}$) where the differences in the circulation patterns of the 1 cm grains and the other smaller grains have also created a double peaked structure in the composite map.}
\label{fig:double_triple_density_maps}
\end{figure*}

\begin{figure*} 
\centering
\includegraphics[width=0.90\textwidth]{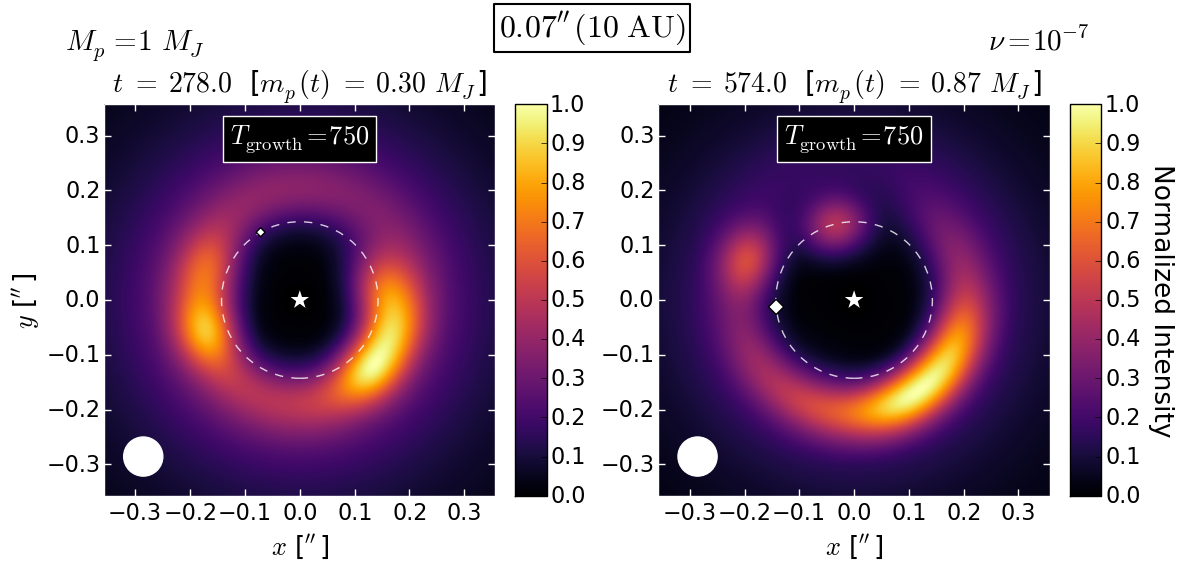} \\
%\vspace*{1.5em}
\includegraphics[width=0.47\textwidth]{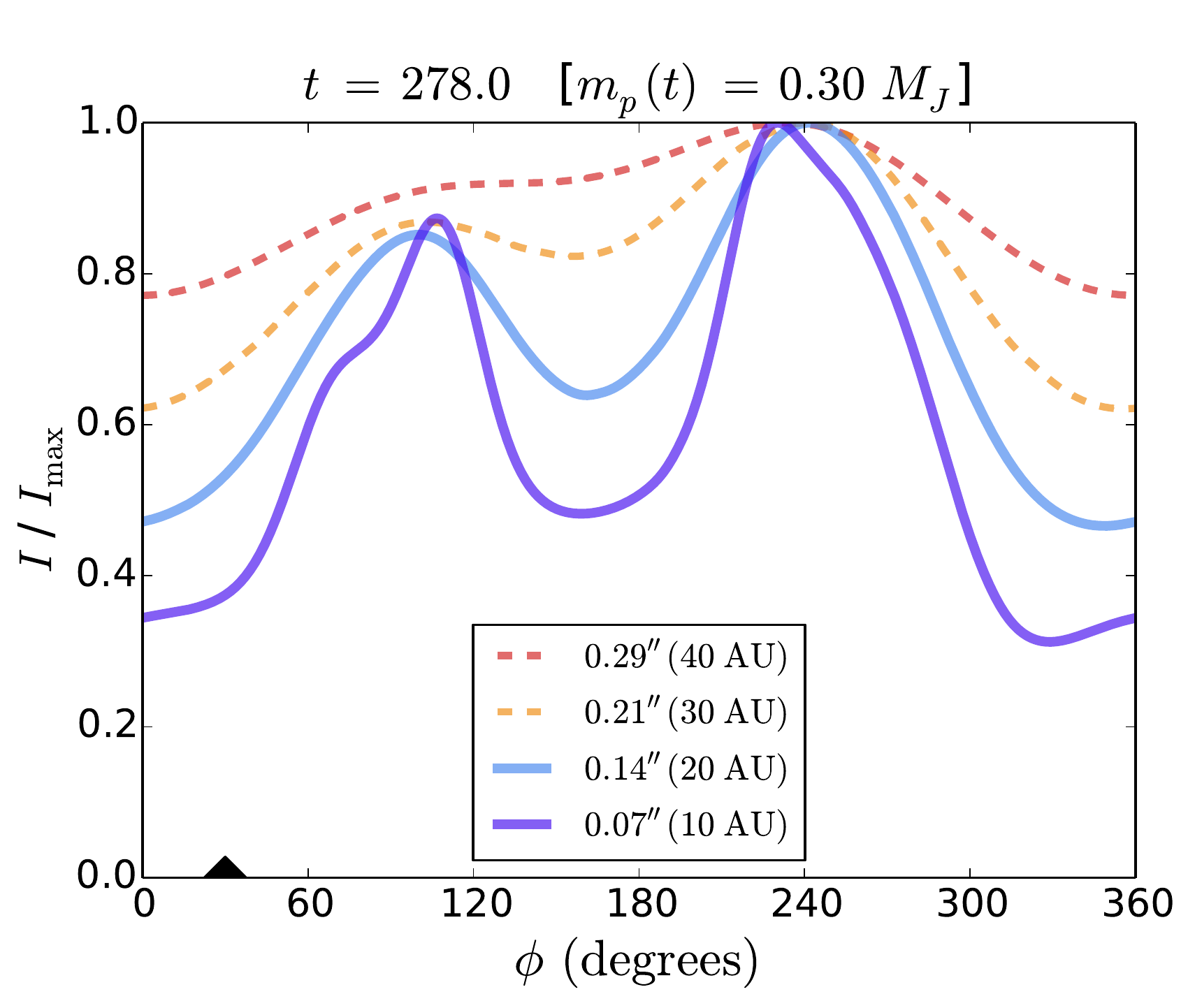}
%\hspace*{0.5em}
\includegraphics[width=0.47\textwidth]{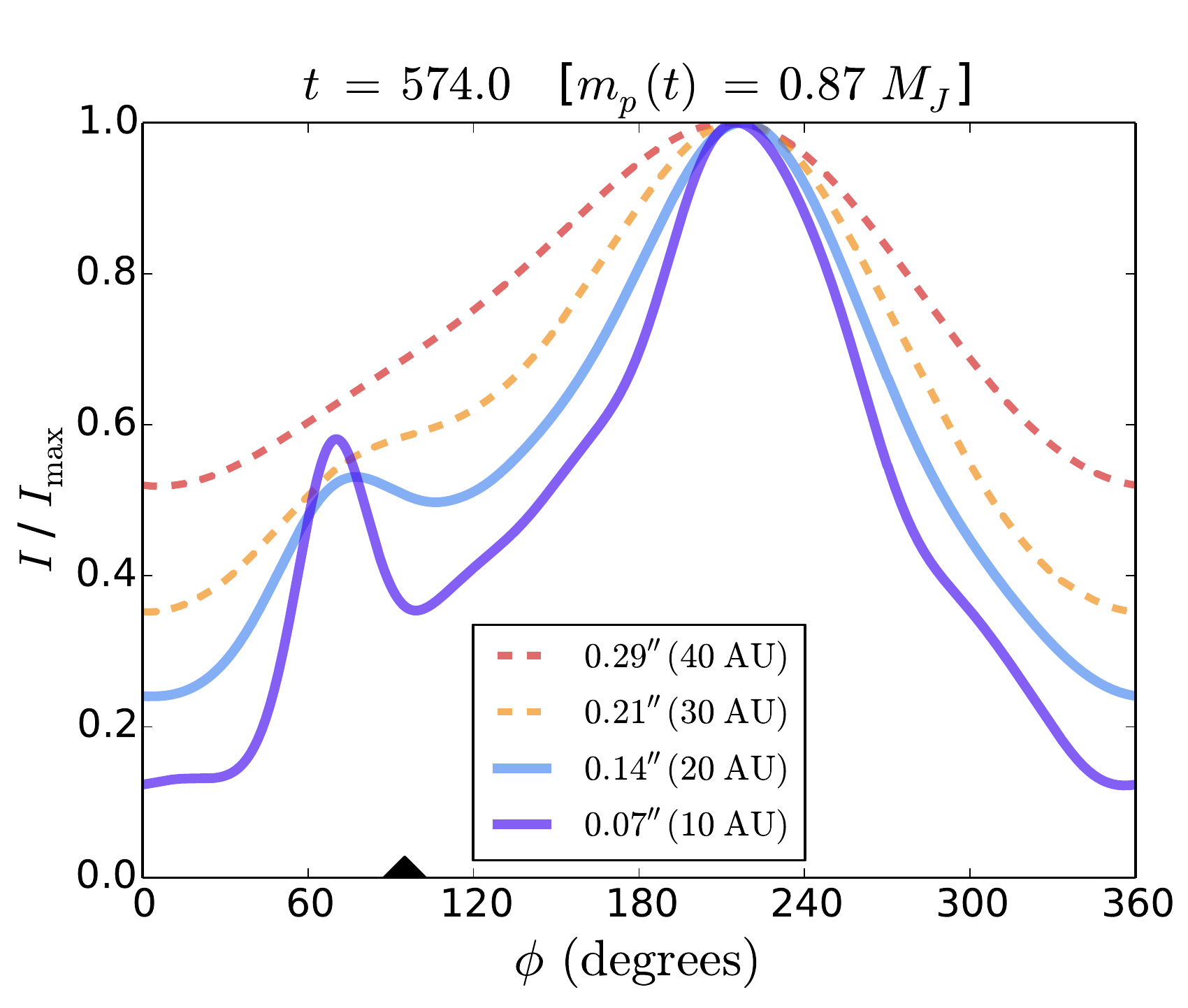}
\caption{Top: Synthetic images of the elongated vortex ($T_\mathrm{growth} = 750$ planet orbits) that exhibit double peaks shortly after the vortex forms ($m_\mathrm{p}(t) = 0.30\ M_\mathrm{Jup}$; left panels) and later on ($m_\mathrm{p}(t) = 0.87\ M_\mathrm{Jup}$; right panels). At the later time, there is also a distinct signature of the co-orbital dust at L5 inside the vortex. Bottom: Comparison of central azimuthal intensity profiles for four different beam sizes. The black triangle indicates the location of the planet. There is a clear double-peak signature with a beam diameter of $<0.14^{\prime \prime}$ in the earlier phase and $<0.07^{\prime \prime}$ in the later phase.}
\label{fig:double_synthetic_images}
\end{figure*}

{\cbf
When a vortex initially forms, it can have a very different double-peaked structure as a signature. This structure arises from the lag between the dust vortices merging relative to the gas vortices. As the higher-number modes merge into a single $m = 1$ mode, they are still spread out in azimuth (see the left panel of Figure~\ref{fig:double_triple_density_maps}). This larger azimuthal extent of the end-product vortex prevents the separate peaks of the previously co-orbital vortices from naturally merging into a single peak on a short timescale, like they do in the concentrated case. Instead, the double-peaked structure persists for over 100 orbits for the 1 cm grains in our fiducial elongated vortex simulation. The smaller 3 mm grains only maintain this pattern for about 60 orbits. 

In some snapshots, the double peaks of different-sized grains are aligned (as shown in the left panels of Figures~\ref{fig:double_triple_density_maps}~and~\ref{fig:double_synthetic_images}), while in others they can be far apart. When these peaks are not aligned, the resulting azimuthal intensity profiles are relatively flat due to the lack of a distinct peak location. It is also possible for the double-peaked signature to arise later in the vortex's lifetime (as shown in the right panels of Figures~\ref{fig:double_triple_density_maps}~and~\ref{fig:double_synthetic_images}); however, one of the peaks is typically dominant over the other in this case. The early-on double-peaked structure resembles the signature found by \cite{baruteau16} for the largest grains in their study of vortices with self-gravity. Multi-wavelength observations should be able to distinguish between these two scenarios, as the shifts in their study have a much steeper monotonic size dependence compared to those in our study, where the shifts of different-sized grains can still overlap and have less of a predictable pattern in general.
}

\subsubsection{Other Features}  \label{sssec:other-features}

%%% ### Outline: New Paragraph 1 ### %%%
%% (6) vortices in the inner disk --- add role of drift

{\cbf
Besides the vortex exterior to the planet, the inner edge of the gap also triggers a vortex interior to the location of the planet. In the case of the elongated vortex, the slow growth of the planet gives the largest dust particles time to drift towards the inner edge of the disk while the planet itself prevents much of the dust in the outer disk from reaching the inner disk. As a result, the interior vortices should appear weaker than the exterior vortices due to the lower supply of the largest dust particles that attain the highest levels of concentration in vortices located in the outer disk.
%We do not discuss these interior vortices in further detail because the arithmetic spacing of the grid in our simulations is not sufficient to properly resolve them. 
}

%%% ### Outline: New Paragraph - ### %%%
%% (-) external traps

%Outside of the primary dust trap, elongated vortices are often accompanied by secondary dust traps slightly exterior to the main asymmetry. {\cbf These external traps appear at local vorticity minima (see right panel of Figure~\ref{fig:vorticity}), which arise due to irregularities in the structure of the gas surrounding the vortices.} The lack of azimuthal symmetry at the gap as well as the planet's spiral density waves may contribute to producing these features. These secondary traps are much weaker, with dust-to-gas ratios about 10 times lower than in the vortex itself. They also do not appear in the simulated images (see Section~\ref{ssec:si-results}) because they do not overlap between the simulations of different-sized dust. However, a real image of a disc that is well-resolved and contains an elongated vortex may have these features.

%%% ### Outline: New Paragraph 7 ### %%%
%% (7) trojan dust

{\cbf
Aside from the vortices that trap dust, the planet also collects dust at the L5 Lagrange point located $60^{\circ}$ behind its location in both cases of short and long growth times. As shown in Figure~\ref{fig:vorticity}, this point also corresponds to a local vorticity minimum. Since the rest of the gap is empty, this co-orbital dust could be visible in observations, as shown in the top panels of Figure~\ref{fig:synthetic_images} and the top right panel of Figure~\ref{fig:double_synthetic_images}. Detecting such a signature would help confirm that a vortex was indeed triggered by a planet and also reveal the planet's location.
}

\section{Applications to Observations} \label{sec:applications}

%%% ### Outline: Entire Section ### %%%
%%(0) Asymmetries are not necessarily caused by planets. Even if it is caused by a planet, it would be difficult to extract properties of planet or disk from the vortex. \\
%%(1) We can distinguish an elongated vortex induced by a planet from one generated by a dead zone boundary through the interaction of the planet's spiral density waves. \\
%%(2) Finding an elongated vortex induced by a planet could show that it may have been triggered while the planet is still growing. Finding a concentrated vortex could show that a planet completes its runaway growth phase relatively quickly, suggesting material is supplied faster than the viscosity (such as with a disk wind). It could also suggest that the planet doesn't trigger a vortex until after it is finished growing. \\
%%(3) With current resolutions, it is difficult to tell if these asymmetries are concentrated or elongated. As a result, models of the system could be inaccurately modeling the mass of the planet or the viscosity of the disk. \\
%%(4) The interaction of the planet's spiral density waves with the vortex makes it possible to observe changes relatively quickly. \\

\subsection{Confirmation of Elongated Planet-Induced Vortices} \label{ssec:confirmation}

To complement our criteria for determining the shape of a vortex, we suggest the following methods for assessing whether an observed asymmetry may be induced by a planet, rather than the boundary of a dead zone:

\begin{enumerate}[label={\arabic{enumi}.}, leftmargin=0.5cm, itemindent=-0.2cm] 
 \item \textbf{Detecting trojan dust co-orbital with the planet:} {\cbf As stated in Section~\ref{sssec:other-features}, an elongated planet-induced vortex should also be accompanied by dust at the L5 Lagrange point -- that is, co-orbital with and $60^{\circ}$ behind the location of the planet. 
 
 \tab \tab \tab \tab Even with small beam sizes ($0.07^{\prime \prime}$), this dust feature should overlap with the interior side of the vortex in the outer disk if it is located in the azimuthal range of the vortex (see the top right panel of Figure~\ref{fig:synthetic_images}). This feature is weaker and could be missing if the vortex is newly formed, which would suggest the planet is in the early stages of its runaway gas growth phase.}

 \item \textbf{Distinguishing from dead zone boundary vortices:} Previous studies of vortices generated by dead zone boundaries have shown that they are also elongated \citep[e.g.][]{regaly12, lyra12}. We suggest two ways to distinguish them:
 
  \tab \tab \tab \tab First, planet-induced vortices frequently have off-center peaks due to repeated perturbations from the planet's spiral arms. On the contrary, \cite{regaly17} find that without these interactions, vortices at dead zone boundaries remain symmetric even when they have wide azimuthal extents.
    
  %On the other hand, symmetric structure could not be used to identify the source of the vortex since both types of vortices may appear symmetric.
 
  \tab \tab \tab \tab Second, \cite{regaly17} find that vortices at dead zone boundaries are much more radially-extended than those induced by rapidly-grown planets. In contrast, we have shown that elongated vortices have {\cbf similar radial extents than concentrated vortices, providing another distinguishing criterion.}
  
 %However, we expect constraining this property to require even higher resolution than what is needed to measure an asymmetry's azimuthal extent since the former is much smaller than the latter.
 
 \item \textbf{Distinguishing from dead zone boundary vortices with feedback:} {\cbf In an alternative model with a longer simulation time and dust feedback, \cite{miranda17} find the dust structure of a dead zone is rather concentrated at the center even though the gas structure is still very elongated.}
 
  \tab \tab \tab \tab {\cbf Furthermore, in their ``GasHigh" run that is the most similar to our study, the primary concentration of dust at the center breaks apart into several smaller clumps due to feedback and a much higher dust-to-gas ratio of $\Sigma_\mathrm{d} / \Sigma_\mathrm{g} \approx 10$. These secondary clumps spread out widely in azimuth, while the primary clump remains dominant (see their Figure 3c). We expect this pattern to produce multiple peaks at a given radial distance that would be easy to distinguish from the singular primary off-center peaks that characterize our synthetic images in all but the initial stages of vortex formation.}
 
 \tab \tab \tab \tab {\cbf It is possible that feedback may also cause elongated planet-induced vortices to break up into clumps. Nonetheless, these multi-clump planet-induced vortices should still be distinguishable from dead zone vortices. In the dead zone vortex model from \cite{miranda17}, there is hardly any dust between separate clumps since the vortex was originally concentrated before it broke apart. This is distinct from our study, in which the dust is always spread out across the entire extent of the vortex. Thus, we expect the presence of dust between adjacent peaks to be unique to elongated planet-induced vortices.}
 
 \item \textbf{Observing time variability in elongated vortices:} A planet's spiral arms can change the dust structure of elongated vortices over timescales as short as one-third of an orbit.
 
   \tab \tab \tab \tab For a planet located at 5 AU around a solar mass star, we may be able to see changes in the azimuthal structure of an observed asymmetry in about 3 to 4 years if it were an elongated vortex -- albeit not always. In our simulations, we only see a dramatic shift in the peak or the azimuthal profiles within a single orbit occurring about once every five orbits. Presently, no system has an asymmetry at a small enough separation to look for variability on these short timescales.
   
\end{enumerate}
   
\subsection{Comparisons to Theoretical Models} \label{ssec:theory}

\begin{enumerate}[label={\arabic{enumi}.}, leftmargin=0.5cm, itemindent=-0.2cm] 
 
 \item \textbf{Modeling planet and disc parameters of vortices:} There is widespread interest in inferring the masses and orbits of unseen planets from the features they create in their discs -- most notably vortices \citep[e.g.][]{bae16}, spiral arms \citep[e.g.][]{dong15}, and gaps \citep[e.g.][]{dipierro17}. It is already difficult to constrain these properties with vortices since the characteristics of a vortex are degenerate between different combinations of planet masses, disc viscosities, and temperature profiles, as well as other factors \citep[e.g.][]{fu14a}. The shape of a vortex adds an additional complication.
 
   \tab \tab \tab \tab The resolutions used to observe systems with asymmetries are not always sufficient to discern whether these features are concentrated or elongated (see Figure~\ref{fig:synthetic_images}). As a result, models that assume these vortices are concentrated may be underestimating the planet's mass or overestimating the disc viscosity, among other possibilities. For these systems, higher resolution images are needed to resolve the shapes of these asymmetries in order to provide better input for predicting the properties of these undetected planet candidates and their discs.
   
 \item \textbf{Validating planet formation models:} There are few good observational methods to constrain planet formation timescales for giant planets. Our best constraints come from measuring the ages of stars in systems with planet candidates this size \citep[e.g.][]{sallum15, donati16, keppler18} or the lifetimes of protoplanetary discs in general \citep[e.g.][]{bell13}, both of which naturally overestimate the ages of the associated planets. The discovery of an elongated planet-induced vortex could offer some more direct insight into the process by which giant planets form.
 
 \tab \tab \tab \tab First, it would show that the vortex was triggered while its accompanying planet was still growing. Additionally, it would help verify that gas giant planets take a relatively long time to complete the runaway gas accretion phase (at least a few hundred orbits for a Jupiter-mass planet and at least 1000 orbits for a 5 Jupiter-mass planet), as expected for the runaway growth phase of core accretion in a low-viscosity disc \citep{lissauer09, hammer17}.
 
  \tab \tab \tab \tab On the contrary, finding a concentrated asymmetry would not necessarily indicate a planet formed rapidly. It might instead suggest that the vortex did not form until after the planet  finished growing, that other factors such as thermodynamic effects shaped the asymmetry \citep{owen17}, or that a planet may not be responsible for the feature in the disc.
  
% [a]~The existence of this type of vortex could suggest that giant planets in the process of forming can accrete gas at rates that are much faster than what is expected from viscous accretion. This may be possible if a disc wind is the dominate mechanism for accretion at the location of the planet in the disc \citep[e.g.][]{bai13}. [b]~Alternatively, it might indicate that the vortex did not form until after the planet finished most of its growth. [c]~Another possibility would be that other factors such as thermodynamic effects play a strong role in shaping planet-induced vortices \citep{owen17}. [d]~And less promisingly for finding planets, it could suggest that the asymmetry may have been sculpted by a close-in companion star \citep{ragusa17} instead of a planet, among other potential sources.
  
\end{enumerate}

\subsection{Vortex Candidates} \label{ssec:candidates}

%%\begin{figure*} 
%%\centering
%%\includegraphics[width=0.45\textwidth]{figures/RYLup_continuum.pdf}
%%\hspace*{2.5em}
%%\includegraphics[width=0.45\textwidth]{figures/SYCha_continuum.pdf}
%%\caption{Candidate elongated vortices: Ry Lupus and SY Chamaeleon (also known as J10563044). Both discs have a highest intensity contour that spans roughly $180^{\circ}$. Left: The additional two spikes within the contour are likely due to the system's high inclination. Right:   Due to the very low resolution of the beam size for this observation, it is difficult to assess how well this system matches our synthetic images.
%%}
%%\label{fig:candidates}
%%\end{figure*}

In this subsection, we compare our synthetic images to observed disks with asymmetries. We find one good candidate system, HD~135344B, for harboring an elongated planet-induced vortex. We also discuss two other systems -- RY Lupus and T4 -- where an elongated vortex cannot be ruled out as an explanation of their asymmetric features. Throughout this section, we presume that the asymmetries are located at a separation of $1.4~r_p$ and that a beam size corresponding to $1.0~r_p$ is needed to resolve each feature.

There are not many good elongated vortex candidates in large part because most systems with asymmetric features have not been observed with sufficient resolution. Moreover, the wide separations of the majority of these features place them beyond the expected locations of giant planets \citep{bowler18}. 

%We do not find any asymmetries that have a high probability of being confirmed as elongated vortices, in large part because most systems with these features have not been observed with sufficient resolution. Moreover, the wide separations of the majority of these features place them beyond the expected birthplaces of giant planets \citep{bowler18}. Nonetheless, we offer insight on systems with asymmetries where an elongated planet-induced vortex cannot be ruled out as an explanation of these features. Throughout this section, we presume that the asymmetries are located at a separation of $1.4~r_p$ and that a beam size corresponding to $1.0~r_p$ is needed to resolve each feature.

\begin{itemize}[leftmargin=0.5cm, itemindent=-0.2cm] 

 \item \textit{HD 135344B} is surrounded by a disk that includes a ring at $0.32^{\prime \prime}$ (45 AU) and an asymmetry at $0.54^{\prime \prime}$ (75 AU), leaving space for a potential planet near $0.38^{\prime \prime}$ (54 AU) in-between the two features \citep{vanDerMarel16b}. It also exhibits spirals in scattered light \citep{muto12, garufi13, stolker16}. Unlike most other systems, its asymmetry has been observed with a small enough beam size to resolve that it has both a wide azimuthal extent a little under $180^{\circ}$ as well as a peak that is noticeably off-center. {\cbf Recent observations have confirmed the presence of a skewness about an off-center peak as well as a decrease in the azimuthal extent at larger wavelengths \citep{cazzoletti18}, suggesting that the asymmetry could be an elongated planet-induced vortex. Without self-gravity in our simulations, we cannot test whether the observed spirals in the system could be caused by the vortex, as \cite{vanDerMarel16b} propose. We recommend future computational studies incorporate self-gravity to study this possibility.}

 %\item \textit{HD 135344B}. Unlike most other systems, HD 135344B has been observed with a small enough beam size to resolve its features \citep{vanDerMarel16b}. The system includes a ring at $0.32^{\prime \prime}$ (45 AU) and an asymmetry at $0.54^{\prime \prime}$ (75 AU), leaving space for a potential planet near $0.38^{\prime \prime}$ (54 AU). It also exhibits spirals in scattered light \citep{muto12, garufi13}. 
 
 %\tab \tab \tab \tab Observations show that the asymmetry indeed has both an elongated extent a little under $180^{\circ}$ as well as a peak that is noticeably off-center.

\end{itemize}
 
\begin{itemize}[leftmargin=0.5cm, itemindent=-0.2cm] 
 
  \item \textit{RY Lupus} has a disk that is inclined at $69^{\circ}$ from the line of sight. It features an asymmetry with an azimuthal extent of roughly $240^{\circ}$ located at $\approx 0.33^{\prime \prime}$ (50 AU) and a contrast of about 1.5 \citep{ansdell16}. Its azimuthal intensity profiles are relatively flat across the asymmetry. The system was observed with a beam size of $0.38^{\prime \prime} \times 0.33^{\prime \prime}$. A beam diameter of at most $0.20^{\prime \prime}$, if not smaller due to the system's high inclination, is needed to better assess the shape of the asymmetry.
  
  %\item \textit{RY Lupus}. The disc around RY Lupus has an elongated asymmetry with an azimuthal extent of roughly $240^{\circ}$ located at $\approx 0.33^{\prime \prime}$ (50 AU) and a contrast of about 1.5 \citep{ansdell16}. Its azimuthal intensity profiles are relatively flat across the asymmetry. The system was observed with a beam size of $0.38^{\prime \prime} \times 0.33^{\prime \prime}$. A beam diameter of at most $0.20^{\prime \prime}$, if not larger due to the system's high inclination, is needed to assess the asymmetry.
  
\end{itemize}
 
\begin{itemize}[leftmargin=0.5cm, itemindent=-0.2cm] 
   
  \item \textit{T4} (also known as \textit{J10563044}) has a relatively face-on disk containing an elongated asymmetry at $\approx 0.40^{\prime \prime}$ (64 AU) with an off-center peak \citep[][]{pascucci16} . The feature has an azimuthal extent of about $160^{\circ}$ and a very low contrast of $\sim 1.2$. With the relatively large beam size of $0.71^{\prime \prime} \times 0.47^{\prime \prime}$, it is difficult to tell if the feature is elongated. A much smaller beam diameter of at most $0.24^{\prime \prime}$ is needed to examine this feature in detail.
    
  %\item \textit{T4} (also known as \textit{J10563044}). Observations of T4 \citep[][]{pascucci16} show it has a peaked elongated asymmetry at $\approx 0.40^{\prime \prime}$ (AU) with an azimuthal extent of about $160^{\circ}$ and a very low contrast of $\sim 1.2$. A relatively large beam size of $0.71^{\prime \prime} \times 0.47^{\prime \prime}$ was used to observe the system. A beam diameter of at most $0.24^{\prime \prime}$ is needed to examine this feature in detail.
 
\end{itemize}

\section{Conclusions} \label{sec:conclusions}

%%% ### Outline: Paragraph 1 ### %%%
%%Outline: \\
%%(0) Overview. \\
%%(1) List three ways to classify elongated vortices. \\
%%(2) Mention how this changes for disks with different masses. \\

We have shown that elongated vortices appear very distinct from concentrated ones in observations taken with beam diameters no larger than the planet's semimajor axis. We suggest  that an asymmetry can be classified as an elongated vortex if it is characterized by the following criteria:

\begin{enumerate}[leftmargin=1cm, itemindent=-0.2cm]
  \item a wide azimuthal extent of at least $180^{\circ}$,
  \item an off-center peak, and 
  {\cbf
  \item a skewness (lack of symmetry) about the peak, or
  \item double peaks (primarily in newly-formed vortices).}
\end{enumerate}

As with standard Kida vortices, observations at different wavelengths can verify that the asymmetry is a dust-trapping vortex in general by showing that the feature is more concentrated at larger wavelengths due to the larger particles concentrating more narrowly in the azimuthal direction \citep{vanDerMarel15b}. {\cbf However, the gap in azimuthal extents of different-sized grains can vary considerably. In rare instances, there may not be a noticeable gap between different sized grains. At later times, smaller-sized grains ($s \leq 1$ mm) converge to a similar extent.} Furthermore, other studies have found this signature may fade away if the gas vortex has already dissipated \citep{fuente17, surville18}.

%Our fiducial runs focused on a single characteristic disc mass, which sets the stopping time of the particles that dominate the synthetic images. If discs are typically more massive, even cm-sized particles cease to have peaked azimuthal profiles in an elongated vortex. Since low-mm wavelength images do not probe much larger grain sizes, we would expect high mass discs to have flat azimuthal intensity profiles at these wavelengths. Thus, observations of high mass discs might not show an elongated vortex's characteristic off-center peaks. Notably, if high mass discs do not possess grains with $s \gg 1$ cm, their vortices might be less subject to dust feedback because the dust-to-gas ratio remains low \citep{fu14b, crnkovic15}.

%%% ### Outline: Paragraph 2 ### %%%
%%Outline: \\
%%(1) List four applications to observations. \\

We offer several suggestions for how to interpret observations of asymmetries that may be elongated vortices. In particular,
\begin{enumerate}[label={\arabic{enumi}.}, leftmargin=0.42cm, itemindent=-0.2cm] 
 \item {\cbf An elongated vortex induced by a planet may also be accompanied by trojan dust that is co-orbital with the planet at the L5 Lagrange point.}
 \item We can distinguish an elongated vortex generated by a planet from one generated at a dead zone boundary by observing a planet-induced vortex's off-center peaks or by resolving the planet-induced vortex's much thinner radial extent.
  \item Another signature of elongated vortices is that the interaction between the planet's spiral density waves and its associated vortex could change the azimuthal intensity profiles of a vortex within a few years for a planet at Jupiter's separation from the Sun, even when the waves are not overlapping the vortex.
 \item If an asymmetry is wide enough to possibly be elongated, resolutions equal to or less than the planet's semimajor axis are needed to model the planet mass that is necessary to explain the asymmetry with a planet-induced vortex. Without this high of a resolution, it would be difficult to classify an asymmetry as concentrated or elongated and in turn make an accurate model of the vortex.
 \item Discovering an elongated planet-induced vortex would help show that gas giant planets generate vortices during their runaway gas accretion phase while they are still forming. The existence of such a vortex would also show that this growth phase takes relatively long to complete (at least several hundred orbits or more, depending on the planet and disc parameters). In contrast, a concentrated asymmetry may favor an alternate explanation.

\end{enumerate}

%%% ### Outline: Paragraph 3 ### %%%
%%Outline: \\
%%(1) Talk about comparisons to real disks. \\

We propose that HD 135344B may harbor an elongated planet-induced vortex, as it contains an asymmetry with a wide azimuthal extent, more concentrated azimuthal extents at longer wavelengths, and an off-center peak. {\cbf However, follow-up modeling that incorporates the system's parameters would be needed to show further support for this idea.} No other observed system has shown favorable signs that it contains an elongated vortex, in part due to the resolution used thus far in ALMA observations. We suggest RY Lupus as another system with an asymmetry that may be a good elongated planet-induced vortex candidate. Confirming this hypothesis would require observations taken with a beam diameter of the planet's assumed separation at $\approx 0.20^{\prime \prime}$, if not higher resolution due to the system's high inclination.

\section*{Acknowledgements}
We thank the referee for many helpful comments that greatly improved this manuscript. We thank Zhaohuan Zhu for providing the two-fluid FARGO code. We would also like to thank Megan Ansdell, Paolo Cazzoletti, and Nienke van der Marel for helpful discussions on the candidate vortex images. Additionally, we thank Jeffrey Fung for discussions of the planet's co-orbital dust. MH is supported by the NSF Graduate Research Fellowship under Grant No. DGE 1143953. PP is supported by NASA through Hubble Fellowship grant HST-HF2-51380.001-A awarded by the Space Telescope Science Institute, which is operated by the Association of Universities for Research in Astronomy, Inc., for NASA, under contract NAS 5-26555. KMK is supported by the National Science Foundation under Grant No. AST-1410174. MKL is supported by the Theoretical Institute for Advanced Research in Astrophysics (TIARA) based in Academica Sinica Institute of Astronomy and Astrophysics (ASIAA). This work is also supported by NASA Astrophysics Theory Program grant NNX17AK59G. The El Gato supercomputer, which is supported by the National Science Foundation under Grant No. 1228509, was used to run all of the simulations in this study.

% #### BIBLIOGRAPHY #### %

%%%%%%%%%%%%%%%%%%%%%%%%%%%%%%%%%%%%%%%%%%%%%%%%%%
%% References

%the-bibliography

% #### APPENDIX: DOUBLE PEAK STRUCTURES #### %

%\newpage

%\newpage

%\appendix{}
%\section{Double-peaked Structures}

%%%%%%%%%%%%%%%%%%%%%%%%%%%%%%%%%%%%%%%%%%%%%%%%%%

\begin{thebibliography}{}

\bibitem[\protect\citeauthoryear{{Ansdell}, {Williams}, {van der Marel},
  {Carpenter}, {Guidi}, {Hogerheijde}, {Mathews}, {Manara}, {Miotello},
  {Natta}, {Oliveira}, {Tazzari}, {Testi}, {van Dishoeck} \& {van
  Terwisga}}{{Ansdell} et~al.}{2016}]{ansdell16}
{Ansdell} M.,  {Williams} J.~P.,  {van der Marel} N.,  {Carpenter} J.~M.,
  {Guidi} G.,  {Hogerheijde} M.,  {Mathews} G.~S.,  {Manara} C.~F.,  {Miotello}
  A.,  {Natta} A.,  {Oliveira} I.,  {Tazzari} M.,  {Testi} L.,  {van Dishoeck}
  E.~F.,    {van Terwisga} S.~E.,  2016, \apj, 828, 46

\bibitem[\protect\citeauthoryear{{Ataiee}, {Pinilla}, {Zsom}, {Dullemond},
  {Dominik} \& {Ghanbari}}{{Ataiee} et~al.}{2013}]{ataiee13}
{Ataiee} S.,  {Pinilla} P.,  {Zsom} A.,  {Dullemond} C.~P.,  {Dominik} C.,
  {Ghanbari} J.,  2013, \aap, 553, L3

\bibitem[\protect\citeauthoryear{{Bae}, {Zhu} \& {Hartmann}}{{Bae}
  et~al.}{2016}]{bae16}
{Bae} J.,  {Zhu} Z.,    {Hartmann} L.,  2016, \apj, 819, 134

\bibitem[\protect\citeauthoryear{{Barenfeld}, {Carpenter}, {Ricci} \&
  {Isella}}{{Barenfeld} et~al.}{2016}]{barenfeld16}
{Barenfeld} S.~A.,  {Carpenter} J.~M.,  {Ricci} L.,    {Isella} A.,  2016,
  \apj, 827, 142

\bibitem[\protect\citeauthoryear{{Barge}, {Ricci}, {Carilli} \&
  {Previn-Ratnasingam}}{{Barge} et~al.}{2017}]{barge17}
{Barge} P.,  {Ricci} L.,  {Carilli} C.~L.,    {Previn-Ratnasingam} R.,  2017,
  \aap, 605, A122

\bibitem[\protect\citeauthoryear{{Barge} \& {Sommeria}}{{Barge} \&
  {Sommeria}}{1995}]{barge95}
{Barge} P.,  {Sommeria} J.,  1995, \aap, 295, L1

\bibitem[\protect\citeauthoryear{{Baruteau} \& {Zhu}}{{Baruteau} \&
  {Zhu}}{2016}]{baruteau16}
{Baruteau} C.,  {Zhu} Z.,  2016, \mnras, 458, 3927

\bibitem[\protect\citeauthoryear{{Bell}, {Naylor}, {Mayne}, {Jeffries} \&
  {Littlefair}}{{Bell} et~al.}{2013}]{bell13}
{Bell} C.~P.~M.,  {Naylor} T.,  {Mayne} N.~J.,  {Jeffries} R.~D.,
  {Littlefair} S.~P.,  2013, \mnras, 434, 806

\bibitem[\protect\citeauthoryear{{Birnstiel}, {Dullemond} \&
  {Brauer}}{{Birnstiel} et~al.}{2010}]{birnstiel10}
{Birnstiel} T.,  {Dullemond} C.~P.,    {Brauer} F.,  2010, \aap, 513, A79

\bibitem[\protect\citeauthoryear{{Bohren} \& {Huffman}}{{Bohren} \&
  {Huffman}}{1983}]{bohren83}
{Bohren} C.~F.,  {Huffman} D.~R.,  1983, {Absorption and scattering of light by
  small particles}

\bibitem[\protect\citeauthoryear{{Bowler} \& {Nielsen}}{{Bowler} \&
  {Nielsen}}{2018}]{bowler18}
{Bowler} B.~P.,  {Nielsen} E.~L.,  2018, ArXiv e-prints

\bibitem[\protect\citeauthoryear{{Casassus}, {van der Plas}, {M}, {Dent},
  {Fomalont}, {Hagelberg}, {Hales}, {Jordan} \& {Mawet} D.}{{Casassus}
  et~al.}{2013}]{casassus13}
{Casassus} S.,  {van der Plas} G.,  {M} S.~P.,  {Dent} W.~R.~F.,  {Fomalont}
  E.,  {Hagelberg} J.,  {Hales} A.,  {Jordan} A.,    {Mawet} D. e.~a.,  2013,
  \nat, 493, 191

\bibitem[\protect\citeauthoryear{{Cazzoletti}, {van Dishoeck}, {Pinilla},
  {Tazzari}, {Facchini}, {van der Marel}, {Benisty}, {Garufi} \&
  {P{\'e}rez}}{{Cazzoletti} et~al.}{2018}]{cazzoletti18}
{Cazzoletti} P.,  {van Dishoeck} E.~F.,  {Pinilla} P.,  {Tazzari} M.,
  {Facchini} S.,  {van der Marel} N.,  {Benisty} M.,  {Garufi} A.,
  {P{\'e}rez} L.,  2018, ArXiv e-prints

\bibitem[\protect\citeauthoryear{{Clarke} \& {Pringle}}{{Clarke} \&
  {Pringle}}{1988}]{clarke88}
{Clarke} C.~J.,  {Pringle} J.~E.,  1988, \mnras, 235, 365

\bibitem[\protect\citeauthoryear{{Crnkovic-Rubsamen}, {Zhu} \&
  {Stone}}{{Crnkovic-Rubsamen} et~al.}{2015}]{crnkovic15}
{Crnkovic-Rubsamen} I.,  {Zhu} Z.,    {Stone} J.~M.,  2015, \mnras, 450, 4285

\bibitem[\protect\citeauthoryear{{de Val-Borro}, {Artymowicz}, {D'Angelo} \&
  {Peplinski}}{{de Val-Borro} et~al.}{2007}]{deValBorro07}
{de Val-Borro} M.,  {Artymowicz} P.,  {D'Angelo} G.,    {Peplinski} A.,  2007,
  \aap, 471, 1043

\bibitem[\protect\citeauthoryear{{de Val-Borro}, {Edgar}, {Artymowicz} \&
  {Ciecielag}}{{de Val-Borro} et~al.}{2006}]{deValBorro06}
{de Val-Borro} M.,  {Edgar} R.~G.,  {Artymowicz} P.,    {Ciecielag} P. e.~a.,
  2006, \mnras, 370, 529

\bibitem[\protect\citeauthoryear{{Dipierro} \& {Laibe}}{{Dipierro} \&
  {Laibe}}{2017}]{dipierro17}
{Dipierro} G.,  {Laibe} G.,  2017, \mnras, 469, 1932

\bibitem[\protect\citeauthoryear{{Donati}, {Moutou}, {Malo}, {Baruteau}, {Yu},
  {H{\'e}brard}, {Hussain}, {Alencar}, {M{\'e}nard}, {Bouvier}, {Petit},
  {Takami}, {Doyon} \& {Cameron}}{{Donati} et~al.}{2016}]{donati16}
{Donati} J.~F.,  {Moutou} C.,  {Malo} L.,  {Baruteau} C.,  {Yu} L.,
  {H{\'e}brard} E.,  {Hussain} G.,  {Alencar} S.,  {M{\'e}nard} F.,  {Bouvier}
  J.,  {Petit} P.,  {Takami} M.,  {Doyon} R.,    {Cameron} A.~C.,  2016, \nat,
  534, 662

\bibitem[\protect\citeauthoryear{{Dong}, {Zhu}, {Rafikov} \& {Stone}}{{Dong}
  et~al.}{2015}]{dong15}
{Dong} R.,  {Zhu} Z.,  {Rafikov} R.~R.,    {Stone} J.~M.,  2015, \apjl, 809, L5

\bibitem[\protect\citeauthoryear{{Dorschner}, {Begemann}, {Henning}, {Jaeger}
  \& {Mutschke}}{{Dorschner} et~al.}{1995}]{dorschner95}
{Dorschner} J.,  {Begemann} B.,  {Henning} T.,  {Jaeger} C.,    {Mutschke} H.,
  1995, \aap, 300, 503

\bibitem[\protect\citeauthoryear{{Fu}, {Li}, {Lubow} \& {Li}}{{Fu}
  et~al.}{2014}]{fu14a}
{Fu} W.,  {Li} H.,  {Lubow} S.,    {Li} S.,  2014, \apjl, 788, L41

\bibitem[\protect\citeauthoryear{{Fu}, {Li}, {Lubow}, {Li} \& {Liang}}{{Fu}
  et~al.}{2014}]{fu14b}
{Fu} W.,  {Li} H.,  {Lubow} S.,  {Li} S.,    {Liang} E.,  2014, \apjl, 795, L39

\bibitem[\protect\citeauthoryear{{Fuente}, {Baruteau}, {Neri}, {Carmona},
  {Ag{\'u}ndez}, {Goicoechea}, {Bachiller}, {Cernicharo} \&
  {Bern{\'e}}}{{Fuente} et~al.}{2017}]{fuente17}
{Fuente} A.,  {Baruteau} C.,  {Neri} R.,  {Carmona} A.,  {Ag{\'u}ndez} M.,
  {Goicoechea} J.~R.,  {Bachiller} R.,  {Cernicharo} J.,    {Bern{\'e}} O.,
  2017, \apjl, 846, L3

\bibitem[\protect\citeauthoryear{{Garufi}, {Quanz}, {Avenhaus}, {Buenzli},
  {Dominik}, {Meru}, {Meyer}, {Pinilla}, {Schmid} \& {Wolf}}{{Garufi}
  et~al.}{2013}]{garufi13}
{Garufi} A.,  {Quanz} S.~P.,  {Avenhaus} H.,  {Buenzli} E.,  {Dominik} C.,
  {Meru} F.,  {Meyer} M.~R.,  {Pinilla} P.,  {Schmid} H.~M.,    {Wolf} S.,
  2013, \aap, 560, A105

\bibitem[\protect\citeauthoryear{{Hammer}, {Kratter} \& {Lin}}{{Hammer}
  et~al.}{2017}]{hammer17}
{Hammer} M.,  {Kratter} K.~M.,    {Lin} M.-K.,  2017, \mnras, 466, 3533

\bibitem[\protect\citeauthoryear{{Jaeger}, {Mutschke}, {Begemann}, {Dorschner}
  \& {Henning}}{{Jaeger} et~al.}{1994}]{jaeger94}
{Jaeger} C.,  {Mutschke} H.,  {Begemann} B.,  {Dorschner} J.,    {Henning} T.,
  1994, \aap, 292, 641

\bibitem[\protect\citeauthoryear{{Johansen}, {Andersen} \&
  {Brandenburg}}{{Johansen} et~al.}{2004}]{johansen04}
{Johansen} A.,  {Andersen} A.~C.,    {Brandenburg} A.,  2004, \aap, 417, 361

\bibitem[\protect\citeauthoryear{{Keppler}, {Benisty}, {Muller}, {Henning},
  {van Boekel} \& {Cantalloube} F.}{{Keppler} et~al.}{2018}]{keppler18}
{Keppler} M.,  {Benisty} M.,  {Muller} A.,  {Henning} T.,  {van Boekel} R.,
  {Cantalloube} F. e.~a.,  2018, ArXiv e-prints

\bibitem[\protect\citeauthoryear{{Kida}}{{Kida}}{1981}]{kida81}
{Kida} S.,  1981, Journal of the Physical Society of Japan, 50, 3517

\bibitem[\protect\citeauthoryear{{Kley}}{{Kley}}{1999}]{kley99}
{Kley} W.,  1999, \mnras, 303, 696

\bibitem[\protect\citeauthoryear{{Li}, {Colgate}, {Wendroff} \& {Liska}}{{Li}
  et~al.}{2001}]{li01}
{Li} H.,  {Colgate} S.~A.,  {Wendroff} B.,    {Liska} R.,  2001, \apj, 551, 874

\bibitem[\protect\citeauthoryear{{Li}, {Finn}, {Lovelace} \& {Colgate}}{{Li}
  et~al.}{2000}]{li00}
{Li} H.,  {Finn} J.~M.,  {Lovelace} R.~V.~E.,    {Colgate} S.~A.,  2000, \apj,
  533, 1023

\bibitem[\protect\citeauthoryear{{Li}, {Li}, {Koller}, {Wendroff}, {Liska},
  {Orban}, {Liang} \& {Lin}}{{Li} et~al.}{2005}]{li05}
{Li} H.,  {Li} S.,  {Koller} J.,  {Wendroff} B.~B.,  {Liska} R.,  {Orban}
  C.~M.,  {Liang} E.~P.~T.,    {Lin} D.~N.~C.,  2005, \apj, 624, 1003

\bibitem[\protect\citeauthoryear{{Lissauer}, {Hubickyj}, {D'Angelo} \&
  {Bodenheimer}}{{Lissauer} et~al.}{2009}]{lissauer09}
{Lissauer} J.~J.,  {Hubickyj} O.,  {D'Angelo} G.,    {Bodenheimer} P.,  2009,
  \icarus, 199, 338

\bibitem[\protect\citeauthoryear{{Lovelace}, {Li}, {Colgate} \&
  {Nelson}}{{Lovelace} et~al.}{1999}]{lovelace99}
{Lovelace} R.~V.~E.,  {Li} H.,  {Colgate} S.~A.,    {Nelson} A.~F.,  1999,
  \apj, 513, 805

\bibitem[\protect\citeauthoryear{{Lovelace} \& {Romanova}}{{Lovelace} \&
  {Romanova}}{2014}]{lovelace14}
{Lovelace} R.~V.~E.,  {Romanova} M.~M.,  2014, Fluid Dynamics Research, 46,
  041401

\bibitem[\protect\citeauthoryear{{Lyra} \& {Lin}}{{Lyra} \&
  {Lin}}{2013}]{lyra13}
{Lyra} W.,  {Lin} M.-K.,  2013, \apj, 775, 17

\bibitem[\protect\citeauthoryear{{Lyra} \& {Mac Low}}{{Lyra} \& {Mac
  Low}}{2012}]{lyra12}
{Lyra} W.,  {Mac Low} M.-M.,  2012, \apj, 756, 62

\bibitem[\protect\citeauthoryear{{Masset}}{{Masset}}{2000}]{FARGO}
{Masset} F.,  2000, \aaps, 141, 165

\bibitem[\protect\citeauthoryear{{Miranda}, {Li}, {Li} \& {Jin}}{{Miranda}
  et~al.}{2017}]{miranda17}
{Miranda} R.,  {Li} H.,  {Li} S.,    {Jin} S.,  2017, \apj, 835, 118

\bibitem[\protect\citeauthoryear{{Muto}, {Grady}, {Hashimoto} \& {Fukagawa}
  M.}{{Muto} et~al.}{2012}]{muto12}
{Muto} T.,  {Grady} C.~A.,  {Hashimoto} J.,    {Fukagawa} M. e.~a.,  2012,
  \apjl, 748, L22

\bibitem[\protect\citeauthoryear{{Owen} \& {Kollmeier}}{{Owen} \&
  {Kollmeier}}{2017}]{owen17}
{Owen} J.~E.,  {Kollmeier} J.~A.,  2017, \mnras, 467, 3379

\bibitem[\protect\citeauthoryear{{Pascucci}, {Testi}, {Herczeg}, {Long},
  {Manara}, {Hendler}, {Mulders}, {Krijt}, {Ciesla}, {Henning}, {Mohanty},
  {Drabek-Maunder}, {Apai}, {Sz{\H u}cs}, {Sacco} \& {Olofsson}}{{Pascucci}
  et~al.}{2016}]{pascucci16}
{Pascucci} I.,  {Testi} L.,  {Herczeg} G.~J.,  {Long} F.,  {Manara} C.~F.,
  {Hendler} N.,  {Mulders} G.~D.,  {Krijt} S.,  {Ciesla} F.,  {Henning} T.,
  {Mohanty} S.,  {Drabek-Maunder} E.,  {Apai} D.,  {Sz{\H u}cs} L.,  {Sacco}
  G.,    {Olofsson} J.,  2016, \apj, 831, 125

\bibitem[\protect\citeauthoryear{{P{\'e}rez}, {Isella}, {Carpenter} \&
  {Chandler}}{{P{\'e}rez} et~al.}{2014}]{perez14}
{P{\'e}rez} L.~M.,  {Isella} A.,  {Carpenter} J.~M.,    {Chandler} C.~J.,
  2014, \apjl, 783, L13

\bibitem[\protect\citeauthoryear{{Pollack}, {Hubickyj}, {Bodenheimer},
  {Lissauer}, {Podolak} \& {Greenzweig}}{{Pollack} et~al.}{1996}]{pollack96}
{Pollack} J.~B.,  {Hubickyj} O.,  {Bodenheimer} P.,  {Lissauer} J.~J.,
  {Podolak} M.,    {Greenzweig} Y.,  1996, \icarus, 124, 62

\bibitem[\protect\citeauthoryear{{Price}, {Cuello}, {Pinte}, {Mentiplay},
  {Casassus}, {Christiaens}, {Kennedy}, {Cuadra}, {Perez M.}, {Marino},
  {Armitage}, {Zurlo}, {Juhasz}, {Ragusa}, {Laibe} \& {Lodato}}{{Price}
  et~al.}{2018}]{price18}
{Price} D.~J.,  {Cuello} N.,  {Pinte} C.,  {Mentiplay} D.,  {Casassus} S.,
  {Christiaens} V.,  {Kennedy} G.~M.,  {Cuadra} J.,  {Perez M.} S.,  {Marino}
  S.,  {Armitage} P.~J.,  {Zurlo} A.,  {Juhasz} A.,  {Ragusa} E.,  {Laibe} G.,
    {Lodato} G.,  2018, ArXiv e-prints

\bibitem[\protect\citeauthoryear{{Ragusa}, {Dipierro}, {Lodato}, {Laibe} \&
  {Price}}{{Ragusa} et~al.}{2017}]{ragusa17}
{Ragusa} E.,  {Dipierro} G.,  {Lodato} G.,  {Laibe} G.,    {Price} D.~J.,
  2017, \mnras, 464, 1449

\bibitem[\protect\citeauthoryear{{Reg{\'a}ly}, {Juh{\'a}sz} \&
  {Neh{\'e}z}}{{Reg{\'a}ly} et~al.}{2017}]{regaly17}
{Reg{\'a}ly} Z.,  {Juh{\'a}sz} A.,    {Neh{\'e}z} D.,  2017, \apj, 851, 89

\bibitem[\protect\citeauthoryear{{Reg{\'a}ly}, {Juh{\'a}sz}, {S{\'a}ndor} \&
  {Dullemond}}{{Reg{\'a}ly} et~al.}{2012}]{regaly12}
{Reg{\'a}ly} Z.,  {Juh{\'a}sz} A.,  {S{\'a}ndor} Z.,    {Dullemond} C.~P.,
  2012, \mnras, 419, 1701

\bibitem[\protect\citeauthoryear{{Sallum}, {Follette}, {Eisner}, {Close},
  {Hinz}, {Kratter}, {Males}, {Skemer}, {Macintosh}, {Tuthill}, {Bailey},
  {Defr{\`e}re}, {Morzinski}, {Rodigas}, {Spalding}, {Vaz} \&
  {Weinberger}}{{Sallum} et~al.}{2015}]{sallum15}
{Sallum} S.,  {Follette} K.~B.,  {Eisner} J.~A.,  {Close} L.~M.,  {Hinz} P.,
  {Kratter} K.,  {Males} J.,  {Skemer} A.,  {Macintosh} B.,  {Tuthill} P.,
  {Bailey} V.,  {Defr{\`e}re} D.,  {Morzinski} K.,  {Rodigas} T.,  {Spalding}
  E.,  {Vaz} A.,    {Weinberger} A.~J.,  2015, \nat, 527, 342

\bibitem[\protect\citeauthoryear{{Shakura} \& {Sunyaev}}{{Shakura} \&
  {Sunyaev}}{1973}]{alpha}
{Shakura} N.~I.,  {Sunyaev} R.~A.,  1973, \aap, 24, 337

\bibitem[\protect\citeauthoryear{{Stolker}, {Dominik}, {Avenhaus}, {Min}, {de
  Boer} \& {Ginski}}{{Stolker} et~al.}{2016}]{stolker16}
{Stolker} T.,  {Dominik} C.,  {Avenhaus} H.,  {Min} M.,  {de Boer} J.,
  {Ginski} C. e.~a.,  2016, \aap, 595, A113

\bibitem[\protect\citeauthoryear{{Surville} \& {Mayer}}{{Surville} \&
  {Mayer}}{2018}]{surville18}
{Surville} C.,  {Mayer} L.,  2018, ArXiv e-prints

\bibitem[\protect\citeauthoryear{{Tanga}, {Babiano}, {Dubrulle} \&
  {Provenzale}}{{Tanga} et~al.}{1996}]{tanga96}
{Tanga} P.,  {Babiano} A.,  {Dubrulle} B.,    {Provenzale} A.,  1996, \icarus,
  121, 158

\bibitem[\protect\citeauthoryear{{van der Marel}, {Cazzoletti}, {Pinilla} \&
  {Garufi}}{{van der Marel} et~al.}{2016}]{vanDerMarel16b}
{van der Marel} N.,  {Cazzoletti} P.,  {Pinilla} P.,    {Garufi} A.,  2016,
  \apj, 832, 178

\bibitem[\protect\citeauthoryear{{van der Marel}, {Pinilla}, {Tobin}, {van
  Kempen}, {Andrews}, {Ricci} \& {Birnstiel}}{{van der Marel}
  et~al.}{2015}]{vanDerMarel15b}
{van der Marel} N.,  {Pinilla} P.,  {Tobin} J.,  {van Kempen} T.,  {Andrews}
  S.,  {Ricci} L.,    {Birnstiel} T.,  2015, \apjl, 810, L7

\bibitem[\protect\citeauthoryear{{van der Marel}, {van Dishoeck}, {Bruderer},
  {Birnstiel}, {Pinilla}, {Dullemond}, {van Kempen}, {Schmalzl}, {Brown},
  {Herczeg}, {Mathews} \& {Geers}}{{van der Marel}
  et~al.}{2013}]{vanDerMarel13}
{van der Marel} N.,  {van Dishoeck} E.~F.,  {Bruderer} S.,  {Birnstiel} T.,
  {Pinilla} P.,  {Dullemond} C.~P.,  {van Kempen} T.~A.,  {Schmalzl} M.,
  {Brown} J.~M.,  {Herczeg} G.~J.,  {Mathews} G.~S.,    {Geers} V.,  2013,
  Science, 340, 1199

\bibitem[\protect\citeauthoryear{{Weidenschilling}}{{Weidenschilling}}{1977}]{weidenschilling77}
{Weidenschilling} S.~J.,  1977, \mnras, 180, 57

\bibitem[\protect\citeauthoryear{{Youdin} \& {Lithwick}}{{Youdin} \&
  {Lithwick}}{2007}]{youdin07}
{Youdin} A.~N.,  {Lithwick} Y.,  2007, \icarus, 192, 588

\bibitem[\protect\citeauthoryear{{Zhu}, {Nelson}, {Dong}, {Espaillat} \&
  {Hartmann}}{{Zhu} et~al.}{2012}]{zhu12}
{Zhu} Z.,  {Nelson} R.~P.,  {Dong} R.,  {Espaillat} C.,    {Hartmann} L.,
  2012, \apj, 755, 6

\end{thebibliography}
 \end{document}